\documentstyle[supertabular,epsf,rotate]{mn}
\newcommand{\f}{\phantom{2}}
\newcommand{\mc}{\multicolumn}

\newcommand{\ltsimeq}{\raisebox{-0.6ex}{$\,\stackrel 
        {\raisebox{-.2ex}{$\textstyle <$}}{\sim}\,$}} 
\newcommand{\gtsimeq}{\raisebox{-0.6ex}{$\,\stackrel 
        {\raisebox{-.2ex}{$\textstyle >$}}{\sim}\,$}} 

\begin{document}

\title[Extremely red galaxy counterparts to 7C radio sources]
{Extremely red galaxy counterparts to 7C radio sources}

\author[Willott et al.]{Chris J.\ Willott$^{1,2}$\footnotemark, Steve
Rawlings$^{1}$ and Katherine M.\ Blundell$^{1}$\\
$^{1}$ Astrophysics, Department of Physics, Keble Road, Oxford, OX1
3RH, U.K. \\ $^{2}$ Instituto de Astrof\'\i sica de Canarias, C/ Via
Lactea s/n, 38200 La Laguna, Tenerife, Spain \\
}

\maketitle

\begin{abstract}

We present $RIJHK$ imaging of seven radio galaxies from the 7C
Redshift Survey (7CRS) which lack strong emission lines and we use
these data to investigate their spectral energy distributions (SEDs)
with models which constrain their redshifts. Six of these seven
galaxies have extremely red colours ($R-K>5.5$) and we find that
almost all of them lie in the redshift range $1< z < 2$. We also
present near-infrared spectroscopy of these galaxies which demonstrate
that their SEDs are not dominated by emission lines, although
tentative lines, consistent with H$\alpha$ at $z=1.45$ and $z=1.61$,
are found in two objects. Although the red colours of the 7CRS
galaxies can formally be explained by stellar populations which are
either very old or young and heavily reddened, independent evidence
favours the former hypothesis.  At $z \sim 1.5$ at least $1/4$ of {\em
powerful} radio jets are triggered in massive ($> L^{\star}$) galaxies
which formed the bulk of their stars several Gyr earlier, that is at
epochs corresponding to redshifts $z \gtsimeq 5$. If a similar
fraction of all $z \sim 1.5$ radio galaxies are old, then
extrapolation of the radio luminosity function shows that, depending
on the radio source lifetimes, between 10 -- 100 \% of the near-IR
selected extremely red object (ERO) population undergo a radio
outburst at epochs corresponding to $1 < z < 2$. An ERO found
serendipitously in the field of one of the 7CRS radio sources appears
to be a radio-quiet analogue of the 7CRS EROs with an emission line
likely to be [OII] at $z=1.20$.  The implication is that some of the
most massive elliptical galaxies formed the bulk of their stars at $z
\gtsimeq 5$ and these objects probably undergo at least two periods of
AGN activity: one at high redshift during which the black hole forms
and another one at an epoch corresponding to $z \sim 1.5$ .

\end{abstract}

\begin{keywords}
galaxies:$\>$active -- galaxies:$\>$evolution -- galaxies:$\>$formation
\end{keywords}

\footnotetext{Email: cjw@astro.ox.ac.uk}

\section{Introduction}

At both low and high redshifts, powerful radio sources are believed to
reside exclusively in giant elliptical galaxies or their progenitors.
Surveys of radio galaxies and radio-loud quasars can therefore be used
to trace the evolution of such massive galaxies over cosmological
timescales. 3C radio galaxies at $z \gtsimeq 0.6$ show extended
continuum and emission line structures aligned along their radio axes
(McCarthy et al. 1987; Chambers et al. 1987). Investigation of the
cause of this so-called `alignment effect' has shown that, while this
is not due to the bulk of the stellar population, a single mechanism,
e.g. recent star-formation or dust-scattered quasar light, cannot
account for all cases (Best et al. 1997). Best et al. and Blundell \&
Rawlings (1999) have pointed out the importance of factoring in source
age to the interpretation of this. However, it is clear that the
strength of the alignment effect decreases with decreasing radio
luminosity (Lacy et al. 1999a) and is weaker in the observed-frame
near-infrared than in the optical (Dunlop \& Peacock 1993; Eales et
al. 1997). Dunlop \& Peacock (1993) found that the $R-K$ colours of
lower power radio galaxies are consistent with evolved stellar
populations. The same is true of the host galaxies of 3C sources at
$z\sim 1$, once the aligned component has been taken into account
(Best et al. 1998).

The reddest $z \sim 1$ radio galaxy in the 3CRR sample of Laing, Riley
\& Longair (1983) is 3C 65. The light from this object has been shown
to be dominated by a very old stellar population ($\sim 4$ Gyr; Lacy
et al. 1995; Stockton, Kellogg \& Ridgway 1995) although a reddened
quasar nucleus seems to make some contribution to its red colour (Lacy
et al. 1995; Simpson, Rawlings \& Lacy 1999).  Two extremely red radio
galaxies at $z \sim 1.5$ have subsequently been discovered in the
follow-up of the faint Leiden Berkeley Deep Survey (LBDS 53W091 ---
Dunlop et al. 1996; LBDS 53W069 --- Dunlop 1999).  Keck spectroscopy
of these galaxies shows that their optical continua are dominated by
very old stellar populations: ages $\gtsimeq 3$ Gyr according to
Spinrad et al. (1997) and Dunlop (1999) although ages $\sim 1.5 ~ \rm
Gyr$ are preferred by Yi et al. (2000). These observations indicate
very high formation redshifts ($z \gtsimeq 5$) for the bulk of stars
in such galaxies. Peacock et al. (1998) showed that such high
formation redshifts are consistent with our current knowledge of the
cosmic power spectrum on $\sim$ Mpc scales.

Near-infrared (near-IR) surveys have discovered a population of
galaxies which are very faint or undetected at optical wavelengths
(e.g. Elston, Reike \& Reike 1988, 1989; Hu \& Ridgway 1994). Galaxies
with observed colours of $R-K \geq 6$ have come to be known as {\em
Extremely Red Objects} (EROs). With the advent of wide-field near-IR
detectors it is now possible to detect large numbers of such
objects. There are generally two ways in which such red optical to
near-IR colours can be achieved; an evolved stellar population at $z >
1$ or a highly reddened, dusty starburst or AGN. Much activity is
currently focussed on trying to understand which of these two types of
object dominate the observed EROs. As yet, there are only a few clear
determinations for individual sources. The ERO HR 10 has been found to
be a very dusty, ultra-luminous infrared galaxy at $z=1.44$ (Dey et
al. 1999). At the time of writing this is the only near-IR selected
ERO which has been shown to have its extreme colours due to dust,
however some of the optical counterparts of sub-mm blank-field surveys
are also found to be extremely red (Smail et al. 1999). In contrast,
there are a couple of EROs which appear to be unreddened evolved
stellar populations at $z \approx 1.5$ (CL0939+4713B - Soifer et
al. 1999; HDFS 223251-603910 - Stiavelli et al. 1999). Near-IR
spectroscopy by Cimatti et al. (1999) has shown the majority of a
sample of nine red galaxies with $R-K>5$ appear to be old stellar
populations at $0.8 < z < 1.6$, with only a couple of cases of
possible dusty starbursts. Studies with the Hubble Space Telescope
(HST) show the majority of EROs to have E/S0 morphologies (Moriondo,
Cimatti \& Daddi 2000; Stiavelli \& Treu 200). The detection of a
strong clustering signal in the ERO population (Daddi et al. 2000a) is
in agreement with the hypothesis that massive evolved galaxies
dominate the population.

The exact nature of extremely red galaxies is an important test of
models of galaxy formation and evolution. A high space density of EROs
which are evolved, elliptical galaxies at intermediate redshifts would
mean high formation redshifts for most massive galaxies. However, if
the EROs are due to ongoing dusty star-formation, possibly in mergers,
it would support certain hierarchical models for galaxy formation. The
likely redshift range of EROs in near-IR samples limited to $K (2.2
\mu{\rm m}) \sim 19$ is $1 \ltsimeq z \ltsimeq 2$. The main reason for
the lower redshift limit is that the continuum below the 4000 \AA~
break passes into the observed $R$-band, while the upper redshift
limit is the result of an upper limit to the intrinsic stellar
luminosity of a galaxy. Note that the radio-selected EROs from the
LBDS survey have a co-moving space density $\sim 100$ times lower than
that of the near-IR selected EROs.

We have been working on securing optical identifications and redshifts
for a complete sample of 76 radio sources selected at a low radio
frequency, namely 151 MHz.  The flux-density limit for this sample ---
the combination of the 7C-I and 7C-II regions of the 7C Redshift
Survey (hereafter referred to as the 7CRS) --- is $S_{151} \geq 0.5$
Jy and the total sky area covered by regions I and II is 0.013 sr.
The flux-density limit is a factor of 25 lower than the 3CRR sample
and a factor of 40 higher than the 2 mJy 1.4 GHz LBDS survey (assuming
a radio spectral index of 0.8). The lower flux-density selection than
the 3CRR sample means we expect little contribution from non-stellar
emission to the optical continua of 7CRS radio galaxies (c.f. Lacy et
al. 1999a). All the sources have been identified at $K$-band and
optical spectroscopy has led to redshifts for 90\% of the sample
(Willott et al., in prep. and Blundell et al., in prep., will present
full details of the 7C-I and 7C-II samples).  The seven sources for
which optical spectroscopy did not secure a redshift are very faint or
invisible in our optical images and have red optical-to-near-IR
colours. Four of them are bona-fide EROs with $R-K>6$, of which the
most extreme example is undetected at $R$-band with $R-K>7.3$. Since
complete redshift information is crucial for deducing the nature and
evolution of radio sources, we have obtained multi-colour photometry
for these objects in order to model their spectral energy
distributions (SEDs) and hence constrain their redshifts. In addition, we
have obtained some low-resolution near-infrared spectroscopy to search
for emission lines and/or breaks in the continuum.

In this paper, we use these observations to constrain the redshifts of
these sources and attempt to determine whether their extremely red
colours are due to old stellar populations or to the effects of dust
reddening on a younger population. In Section 2 we list the objects
studied in this paper and discuss their colours in the context of the
sources with spectroscopic redshifts in the 7CRS. In Section 3 we
present our observations and in Section 4 we describe the SED fitting
technique used. A discussion of the SED fitting for individual objects
follows in Section 5. In Section 6 we review the reliability of the
redshift estimation. In Section 7 we discuss the implications of our
findings with respect to radio galaxies, the field ERO population and
the galaxy population in general. We report on the serendipitous
discovery of a radio-quiet ERO with a spectroscopic redshift of
$z=1.200$ in the appendix. We assume that $H_0=50~ {\rm
km~s^{-1}Mpc^{-1}}$ and consider two spatially-flat cosmologies:
$\Omega_{\mathrm M} =1$, $\Omega_ \Lambda=0$ and $\Omega_{\mathrm
M}=0.3$, $\Omega_\Lambda=0.7$.

\begin{figure*}
\epsfxsize=0.95\textwidth \epsfbox{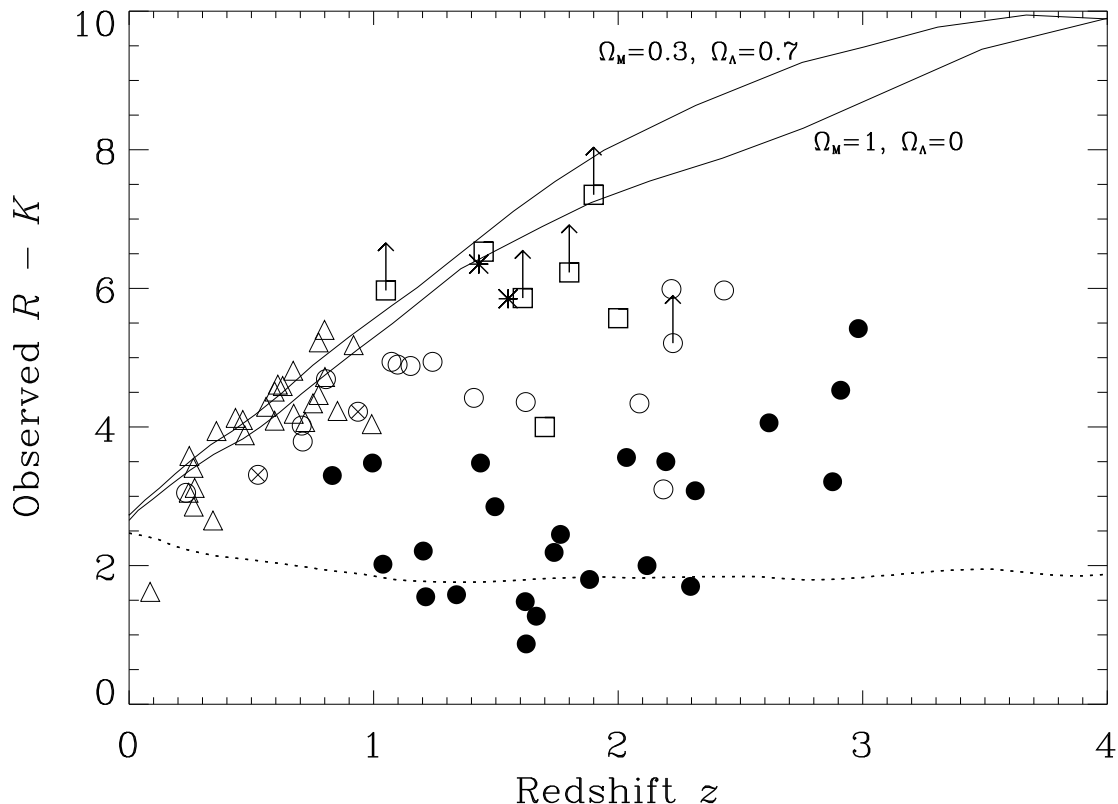}
{\caption[junk]{\label{fig:rkz} Observed $R-K$ colour as a function of
redshift for sources in the 7C-I and 7C-II regions of the 7C Redshift
Survey (7CRS). Filled circles are quasars, crossed circles are
broad-lined radio galaxies, open circles are high emission line
luminosity radio galaxies and open triangles are low emission line
luminosity radio galaxies (see Willott et al. 1998, 2000a for
definitions of these categories). The open squares are the seven 7CRS
objects without redshifts from optical spectroscopy which have
redshifts estimated in this paper. The asterisks show the two red
radio galaxies from the LBDS survey which have redshifts inferred from
stellar absorption lines (Dunlop 1999).  The solid lines show the
evolution of the expected observed colour of a galaxy which formed in
an instantaneous starburst at redshift $z=10$ for $\Omega_{\mathrm M}
=1$, $\Omega_ \Lambda=0$ and for $\Omega_{\mathrm M}=0.3$,
$\Omega_\Lambda=0.7$ using the models of Bruzual \& Charlot
(in prep.). The dotted line shows the change in $R-K$ as a function of $z$
for a typical unreddened quasar spectrum.  }}
\end{figure*}

\section{The sample} 
\label{sample}

In this paper we investigate the nature of the seven radio sources
without redshifts from optical spectroscopy in the 7C-I and 7C-II
regions (based on the 5C6 and 5C7 regions, respectively, of Pearson \&
Kus 1978). Although the flux-density limit of the survey is based on
the 7C survey at 151 MHz, all seven sources without redshifts also
happen to be catalogued 5C sources. They are:

\begin{tabular}{lc}
5C6.17  & (02 06 22.08 +34 14 25.2)\\ 
5C6.62  & (02 10 29.70 +32 54 02.7)\\ 
5C6.83  & (02 11 11.23 +30 39 50.5)\\ 
5C6.242 & (02 18 17.51 +31 03 24.5)\\ 
5C7.47  & (08 12 38.81 +24 56 13.8)\\ 
5C7.208 & (08 20 18.77 +25 06 21.6)\\ 
5C7.245 & (08 22 56.04 +26 53 48.2)\\
\end{tabular}

\noindent where the positions in brackets are epoch B1950.0 and refer to the
location of the $K$-band identification. Radio positions from the 7C
survey and our own VLA maps will be presented in Blundell et al. (in
prep.). All these sources have been observed at low spectral
resolution with the ISIS spectrograph at the WHT with exposures of
between 30 and 60 minutes. Only for 5C7.47 was weak continuum
detected; in all other cases no lines or continuum were detected.
Details of this optical spectroscopy will be given in Willott et
al. (in prep.).

In addition, we include in this study one further source with red
optical to near-IR colours, whose redshift was one of the final ones
to be secured from our optical spectroscopy runs. This object, 5C7.271
(08 26 00.89 +25 04 01.8, B1950.0) has a redshift $z=2.224$ (Willott
et al., in prep.).

\begin{table*}
\footnotesize
\begin{center}
\begin{tabular}{lllcrcll}
\hline\hline
\mc{1}{l}{Name} &\mc{1}{c}{Telescope+} &\mc{1}{c}{Date}&\mc{1}{c}{Filter}&\mc{1}{c}{Exposure} &\mc{1}{c}{Seeing}   &\mc{1}{c}{Magnitude}    &\mc{1}{c}{Magnitude}     \\
\mc{1}{l}{ }    &\mc{1}{c}{Instrument} &\mc{1}{c}{}    &\mc{1}{c}{}      & \mc{1}{c}{time (s)}&\mc{1}{c}{(arcsec)} &\mc{1}{c}{3'' aperture} &\mc{1}{c}{5'' aperture}  \\
\hline\hline
            &  WHT+AUX    & 95Jan25 & $R$ &  300 & 0.7 &  $>$24.5          &  $>$24.0           \\
            &  INT+WFC    & 98Jul27 & $I$ & 1200 & 1.0 &  22.66 $\pm$ 0.15 &  22.11 $\pm$ 0.16  \\
 5C6.17     &  UKIRT+IRC3 & 97Jan29 & $J$ & 1080 & 0.8 &  20.30 $\pm$ 0.11 &  20.10 $\pm$ 0.15  \\
            &  UKIRT+IRC3 & 96Jul29 & $H$ &  540 & 1.1 &  19.09 $\pm$ 0.13 &  18.98 $\pm$ 0.20  \\
            &  UKIRT+IRC3 & 96Feb13 & $K$ & 1440 & 1.4 &  18.41 $\pm$ 0.07 &  18.03 $\pm$ 0.09  \\ 
\hline
            &  MCD+IGI    & 99Oct10 & $R$ & 1800 & 1.7 &  24.57 $\pm$ 0.45 &  $>$24.3           \\
            &  NOT+BRO    & 96Jul23 & $I$ & 1800 & 1.3 &  23.09 $\pm$ 0.38 &  22.76 $\pm$ 0.45  \\
 5C6.62     &  UKIRT+IRC3 & 96Jan21 & $J$ & 1080 & 1.4 &  20.69 $\pm$ 0.18 &  20.25 $\pm$ 0.21  \\
            &  UKIRT+IRC3 & 97Jan30 & $H$ &  480 & 1.1 &  19.01 $\pm$ 0.10 &  18.51 $\pm$ 0.11  \\
            &  UKIRT+IRC3 & 96Jul21 & $K$ &  540 & 1.0 &  18.04 $\pm$ 0.08 &  17.53 $\pm$ 0.09  \\
\hline
            &  WHT+AUX    & 95Jan31 & $R$ &  300 & 0.6 &  $>$24.5          &  $>$24.0           \\
            &  INT+WFC    & 98Jul27 & $I$ & 1200 & 1.5 &  $>$23.8          &  $>$23.3           \\
 5C6.83     &  UKIRT+IRC3 & 97Jan29 & $J$ & 1080 & 1.2 &  20.24 $\pm$ 0.11 &  19.97 $\pm$ 0.15  \\
            &  UKIRT+IRC3 & 97Jan29 & $H$ & 1080 & 1.1 &  19.21 $\pm$ 0.08 &  18.94 $\pm$ 0.10  \\
            &  UKIRT+IRC3 & 97Feb29 & $K$ & 1080 & 1.2 &  18.27 $\pm$ 0.08 &  18.16 $\pm$ 0.13  \\
\hline
            &  WHT+PFC    & 95Nov11 & $R$ & 1800 & 1.6 &  $>$26.1          &  $>$25.6           \\
            &  KECKII+LRIS& 98Oct12 & $I$ &  300 & 0.6 &  $>$25.2          &  $>$24.7           \\
 5C6.242    &  UKIRT+IRC3 & 97Jan29 & $J$ & 1080 & 1.1 &  20.96 $\pm$ 0.25 &  ---               \\
            &  UKIRT+IRC3 & 96Oct29 & $H$ & 1620 & 1.7 &  19.84 $\pm$ 0.40 &  ---               \\
            &  KECK+NIRC  & 98Oct12 & $K$ &  540 & 0.6 &  18.75 $\pm$ 0.10 &  18.58 $\pm$ 0.10  \\
\hline
            &  WHT+AUX    & 95Jan31 & $R$ &  900 & 0.7 &  23.49 $\pm$ 0.12 &  23.72 $\pm$ 0.28  \\
 5C7.47     &  UKIRT+IRC3 & 97Jan29 & $J$ & 1080 & 0.8 &  21.22 $\pm$ 0.30 &  ----              \\
            &  UKIRT+IRC3 & 97Jan29 & $H$ & 1080 & 0.8 &  19.96 $\pm$ 0.20 &  19.99 $\pm$ 0.25  \\
            &  UKIRT+IRC3 & 96Jan21 & $K$ & 4320 & 1.0 &  19.49 $\pm$ 0.10 &  19.34 $\pm$ 0.15  \\
\hline
            &  WHT+AUX    & 95Jan31 & $R$ &  300 & 0.7 &  23.65 $\pm$ 0.21 &  ---               \\
            &  WHT+AUX    & 97Apr07 & $I$ &  900 & 0.7 &  23.26 $\pm$ 0.23 &  ---               \\
 5C7.208    &  UKIRT+IRC3 & 97Jan29 & $J$ &  540 & 1.2 &  20.28 $\pm$ 0.16 &  19.63 $\pm$ 0.15  \\
            &  UKIRT+IRC3 & 97Jan29 & $H$ &  540 & 1.1 &  19.17 $\pm$ 0.13 &  18.83 $\pm$ 0.17  \\
            &  UKIRT+IRC3 & 95Mar01 & $K$ & 1620 & 1.1 &  18.08 $\pm$ 0.08 &  17.68 $\pm$ 0.06  \\
\hline
            &  MCD+IGI    & 95Nov19 & $R$ & 1500 & 3.2 &  $>$25.0          &  $>$24.5           \\
            &  WHT+AUX    & 97Apr07 & $I$ &  900 & 0.7 &  $>$24.3          &  $>$23.7           \\
 5C7.245    &  UKIRT+IRC3 & 97Jan29 & $J$ & 1080 & 0.9 &  21.01 $\pm$ 0.18 &  20.69 $\pm$ 0.23  \\
            &  UKIRT+IRC3 & 97Jan29 & $H$ & 1080 & 0.9 &  19.94 $\pm$ 0.13 &  19.50 $\pm$ 0.14  \\
            &  UKIRT+IRC3 & 95Mar01 & $K$ & 2700 & 1.2 &  19.14 $\pm$ 0.08 &  18.73 $\pm$ 0.10  \\
\hline
            &  MCD+IGI    & 95Nov19 & $R$ &  900 & 2.2 &  $>$24.5          &  $>$24.0           \\
            &  WHT+AUX    & 97Apr08 & $I$ &  600 & 0.7 &  $>$24.0          &  $>$23.5           \\
 5C7.271    &  UKIRT+IRC3 & 97Jan29 & $J$ & 1080 & 0.8 &  21.02 $\pm$ 0.18 &  21.15 $\pm$ 0.34  \\
            &  UKIRT+IRC3 & 97Jan29 & $H$ & 1080 & 0.8 &  19.78 $\pm$ 0.13 &  19.54 $\pm$ 0.18  \\
            &  UKIRT+IRC3 & 96Jan21 & $K$ & 1620 & 1.1 &  18.94 $\pm$ 0.09 &  18.79 $\pm$ 0.13  \\
\hline\hline
\end{tabular}
\end{center}              
{\caption[Table of observations]{\label{tab:obsimtab} Log of imaging
observations of the radio galaxies. Magnitudes and limits have been
corrected for galactic extinction as in Willott et al. (1998).
Telescope and instruments are WHT+AUX -- William Herschel Telescope,
Auxiliary-port imager; INT+WFC -- Isaac Newton Telescope, Wide Field
Camera; UKIRT+IRC3 -- United Kingdom InfraRed Telescope, IRCAM3
detector; MCD+IGI -- McDonald Observatory Texas 2.7-m Telescope, IGI
imaging spectrograph; NOT+BRO -- Nordic Optical Telescope, Brorfelde
CCD; WHT+PFC -- WHT, Prime Focus camera; KECKII+LRIS -- Keck-II
Telescope, Low-Resolution Imaging Spectrograph; KECK+NIRC -- Keck
Telescope, Near-InfraRed Camera.}}

\normalsize
\end{table*}

Before we discuss these objects in detail it is instructive to
consider the optical/near-IR properties of the 7CRS. In Fig.
\ref{fig:rkz} we show the $R-K$ colour versus redshift diagram for the
7C-I and 7C-II regions of the 7CRS. The quasars are well-separated
from the radio galaxies on this diagram and in general have values
similar to those expected by redshifting the optically-selected LBQS
composite spectrum of Francis et al. (1991). Although there is scatter
in the quasar colours it is found that the reddest quasars are those
at the highest redshifts. This will be discussed in a future
paper. Radio galaxies on the plot have been divided into two groups
depending upon their emission line luminosities, as in Willott et
al. (2000a). Those with narrow [OII] emission line luminosities
$\log_{10} (L_{\rm [OII]} / {\rm W}) > 35.1$ are shown as open
circles, while those with lower emission line luminosities are shown
as open triangles. The solid curves are models featuring an
instantaneous starburst at redshift $z=10$ for two cosmological
models. At redshifts $z \leq 1$ the colours of the radio galaxies fall
nicely on the model curves suggesting that these colours result from
very old galaxy populations with little (unobscured) current
star-formation. In particular there is an excellent match between the
upper envelope of the observed $R-K$ colours and the $\Omega_M=0.3$,
$\Omega_\Lambda=0.7$ model. Deviations to bluer colours can be caused
by a small amount of more recent star-formation.

Moving to $z \gtsimeq 1$ we find a marked increase in the scatter of
the $R-K$ colours of the radio galaxies. This is probably due to a
combination of two effects. First, at redshifts higher than $z \sim
0.8$, the observed $R$-band samples the rest-frame light below the
4000 \AA~ break. The ratio of fluxes below and above 4000 \AA~ is a
very strong function of the amount of current or recent
star-formation. Therefore small differences in the amount of recent
star-formation will have a more dramatic effect on the observed colour
at $z > 0.8$.  Lilly \& Longair (1984) showed that the optical/near-IR
colours of 3CR galaxies at $z>1$ are inconsistent with a no-evolution
model. Their observed $R-K \approx 4$ colours are similar to those of
the bluest 7CRS radio galaxies. For the 3C objects these colours are
best explained by recent star-formation in a few cases (Chambers \&
McCarthy 1990), but in most cases, as further evidenced by large
optical polarizations (e.g. di Serego Alighieri et al. 1989; Tadhunter
et al. 1992), they are probably caused by an extra non-stellar
rest-frame UV component, typically scattered light from the quasar
nucleus (see Best et al. 1998).

\begin{figure*}
\epsfxsize=0.95\textwidth 
\epsfbox{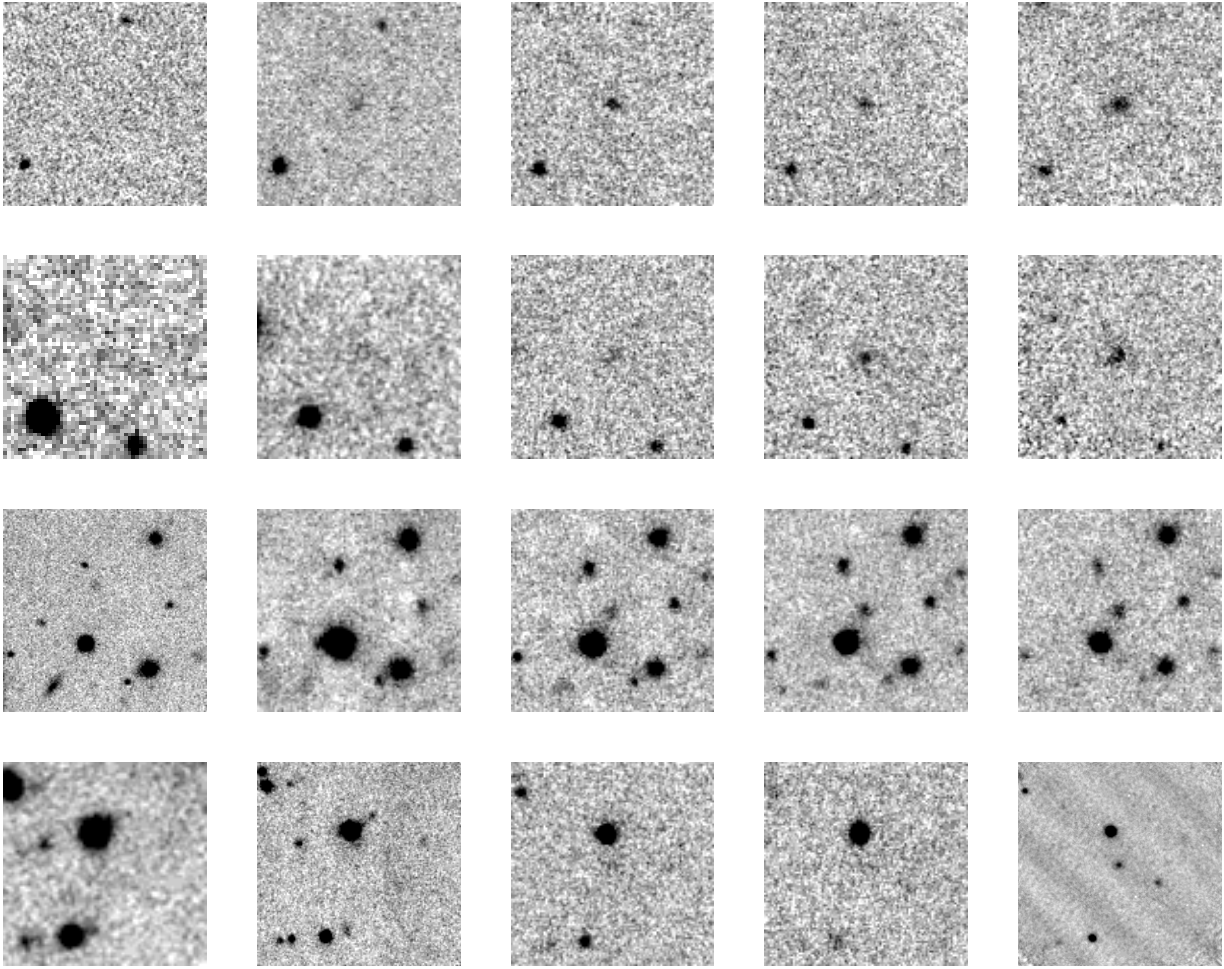}
{\caption[junk]{\label{fig:ims5c6}Central $30 \times 30$ arcsec$^2$ of
the 7C--I (5C6) radio galaxy fields. 
From left to right, $R$, $I$, $J$, $H$ and $K$.
From top to bottom, 5C6.17, 5C6.62, 5C6.83, 5C6.242.}}
\end{figure*}

\begin{figure*}
\epsfxsize=0.95\textwidth 
\epsfbox{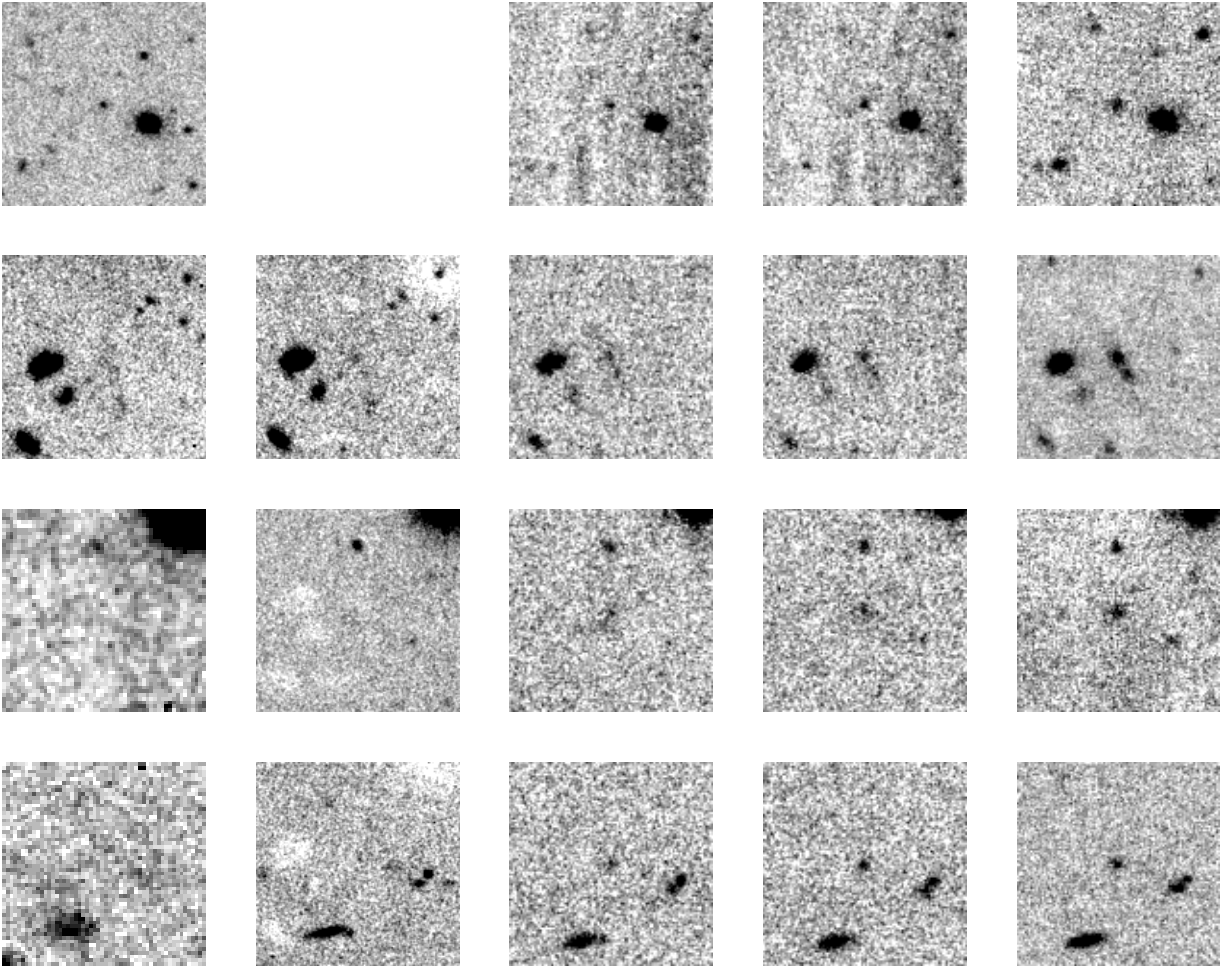}
{\caption[junk]{\label{fig:ims5c7}Central $30 \times 30$ arcsec$^2$ of
the 7C--II (5C7) radio galaxy fields. From left to right, $R$, $I$,
$J$, $H$ and $K$. From top to bottom, 5C7.47 (no $I$ image), 5C7.208,
5C7.245, 5C7.271.}}
\end{figure*}

It is clear from Fig. \ref{fig:rkz} that the bluer 7CRS galaxies are
much more likely to have redshifts measured from optical spectroscopy.
There is also a tendency for the galaxies with high emission line
luminosities (circles) to be bluer than those with low emission line
luminosities (triangles). These observations show that the bluer
sources have stronger emission lines and are therefore good candidates
for having non-stellar (AGN) and/or starburst components in the
rest-frame UV [a similar result was found by Dunlop \& Peacock 1993
for radio galaxies from the Parkes Selected Regions (PSR)]. The large
scatter observed for the high-$z$ 7CRS radio galaxies is however,
quite different from the small scatter in this diagram for PSR
galaxies. With colours $4< R-K <5$ in all cases at $z>0.7$, the PSR
galaxies are similar to the bluest galaxies in the 7CRS, although this
could be partly due to selection effects, since in 1993 only a
fraction of the PSR sample had spectroscopic redshifts (and these were
most likely to be biased in favour of the bluer objects, as is seen
for the 7CRS).  At $z > 1$, radio galaxies from the 6CE sample, which
has a flux-limit in between those of 3CRR and the 7CRS, also tend to
have strong emission lines and blue colours (Rawlings, Eales \& Lacy
2000).

Note from Fig. \ref{fig:rkz} that the reddest 7CRS EROs are not
necessarily intrinsically redder (in the rest-frame) than the
low-redshift radio galaxies.  It is the effect of the steep spectrum
of an evolved galaxy below the 4000 \AA~ break passing through the
$R$-band which has a marked effect on the observed colour at $z
\gtsimeq 0.5$. There are two 7CRS radio galaxies which are at $z>2$
and have very red ($R-K \approx 6$) colours. These both have
spectroscopic redshifts, have been imaged with the Hubble Space
Telescope and are the subject of a future paper.

\section{Observations} 

\subsection{Optical and near-infrared imaging} 

Near-infrared imaging observations in the $J$, $H$ and $K$ bands were
made for all 8 members of the sample studied here using the IRCAM3
detector on the United Kingdom InfraRed Telescope (UKIRT). A deeper
$K$-band image of 5C6.242 has since been obtained from the Keck
telescope equipped with NIRC (thanks to C. Steidel). Table
\ref{tab:obsimtab} details these observations. The near-IR images were
generally taken in multiples of 9 minutes, with a minute at each
position of a 9-point mosaic with typical offsets of 10 arcsec. This
method provides good flat-fielding and sky-subtraction and enables
cosmic rays and bad pixels to be easily removed. The basic data
reduction steps were (i) subtract dark frame; (ii) flat-field using a
flat created from median--filtered object images; (iii) mosaic and
combine the individual object images correcting for atmospheric
extinction, using a bad pixel mask and cosmic ray rejection; (iv) fit
astrometric solution using stars from HST Guide Star Catalogue or APM
catalogue; (v) aperture photometry subtracting off the background. In
all cases (except 5C7.208 -- see Section \ref{notes}) the radio source
counterpart could unambiguously be identified in the $K$-band image by
comparison with the radio images in Blundell et al. (in prep.).

All the near-IR imaging was performed in photometric conditions apart
from the $H$-band image of 5C6.242. From photometric standards
observed before and after this target we estimate that there was 0.8
mag of extinction in $H$-band; during the course of the 27 minute
exposure the magnitudes of bright objects in the field varied by 0.6
magnitudes. We therefore assign a large uncertainty of $\pm 0.4$ mag
to the $H$-band magnitude of this object.

Optical ($R$- and $I$-band) images were obtained from a range of
telescopes and instruments as detailed in Table \ref{tab:obsimtab}.
5C7.47 was not observed in $I$-band. The data were reduced using
standard procedures. The steps involved were (i) subtract bias frame;
(ii) flat-field using twilight sky or dome flats; (iii) correct for
atmospheric extinction; (iv) subtract sky background (including
fringing for some $I$-band images); (v) combine individual frames
(when appropriate), rejecting cosmic rays; (vi) astrometry; (vii)
aperture photometry. For about half the targets no reliable
identification at the position of the near-IR counterpart was found in
either or both the optical images. In these cases we have adopted a
2$\sigma$ upper limit on their fluxes as indicated by the $>$ symbol
in Table \ref{tab:obsimtab}. The optical and near-IR images are shown
in Figs. \ref{fig:ims5c6} and \ref{fig:ims5c7}.

\subsection{Infrared spectroscopy} 

\begin{figure*}
\epsfxsize=0.95\textwidth \epsfbox{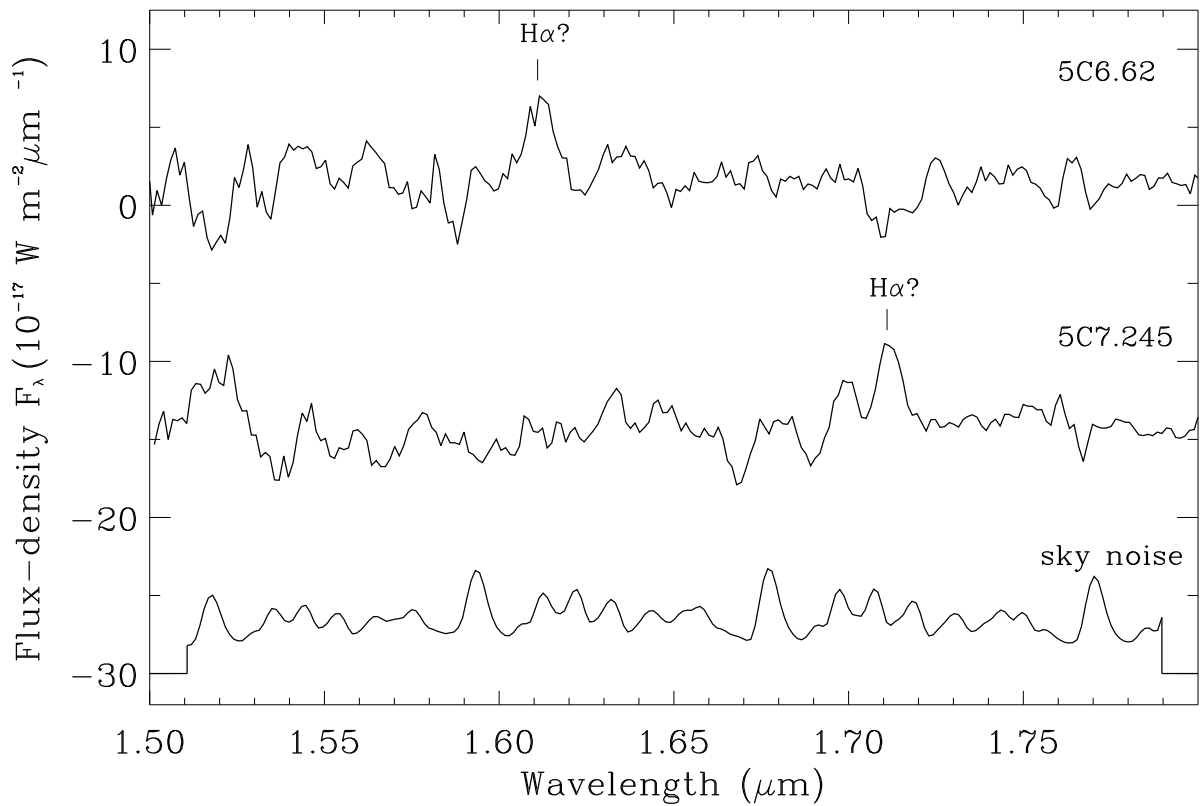}
{\caption[junk]{\label{fig:specgs4} $H$-band spectra of the two radio
sources which show probable ($3\sigma$) emission lines in their CGS4
spectra. The spectrum of 5C7.245 has been displaced by $-15 \times
10^{-17} ~ \rm W \, m^{-2} \, \mu m^{-1}$.  The bottom spectrum shows
the sky noise spectrum (square root of the sky spectrum) displaced by $-30
\times 10^{-17} ~ \rm W \, m^{-2} \, \mu m^{-1}$. Note that the two
probable emission lines are not in regions of high sky noise.  }}
\end{figure*}

\begin{table*}
\footnotesize
\begin{center}
\begin{tabular}{lcccccc}
\hline\hline

\mc{1}{l}{Name} &\mc{1}{c}{Instrumental} &\mc{1}{c}{Wavelength}        &\mc{1}{l}{\f Date} &\mc{1}{c}{Exposure} &\mc{1}{c}{Slit width}  &\mc{1}{c}{Line flux limit} \\

\mc{1}{l}{ }    &\mc{1}{c}{Setup}        &\mc{1}{c}{Coverage ($\mu$m)} &\mc{1}{c}{}     &\mc{1}{c}{time (s)} &\mc{1}{c}{ (arcsec)}  &\mc{1}{c}{(W m$^{-2}$)}  \\
\hline\hline
 5C6.17     &  40 l mm$^{-1}$ + L & 1.5 --- 2.1 &  97Dec30 \f \f  &  2400  & 2.4  &  $2.5 \times 10^{-19}$    \\
 5C6.62     &  40 l mm$^{-1}$ + L & 1.5 --- 2.1 &  97Dec29 \f \f  &  3360  & 4.8  &  $5.0 \times 10^{-19}$    \\
 \f \f \f''     &  40 l mm$^{-1}$ + L & 1.9 --- 2.5 &  97Dec13 \f \f  &  2400  & 4.8  &  $5.5 \times 10^{-19}$    \\
 5C6.83     &  40 l mm$^{-1}$ + L & 1.5 --- 2.1 &  97Dec11/30     &  4480  & 4.8  &  $3.5 \times 10^{-19}$    \\
 5C7.208    &  75 l mm$^{-1}$ + S & 1.5 --- 2.1 &  97Jan30 \f \f  &  1600  & 2.4  &  $3.5 \times 10^{-19}$    \\
 \f \f \f''    &  40 l mm$^{-1}$ + L & 1.9 --- 2.5 &  97Dec12 \f \f  &  1320  & 4.8  &  $6.0 \times 10^{-19}$    \\
 5C7.245    &  75 l mm$^{-1}$ + S & 1.5 --- 2.1 &  97Mar13 \f \f  &  2160  & 2.4  &  $4.5 \times 10^{-19}$    \\
 \f \f \f'' & 75 l mm$^{-1}$ + S & 1.9 --- 2.4 & 97Mar13 \f \f & 1600
& 2.4 & $3.5 \times 10^{-19}$ \\ 
\hline\hline 
\end{tabular}
\end{center} 

{\caption[Table of observations]{\label{tab:cgs4} Log of CGS4
spectroscopy of sources discussed in this paper. The instrumental
setup gives the grating used, either 40 lines mm$^{-1}$ or 75 lines
mm$^{-1}$, and the Long (300 mm) or Short (150 mm) camera. Exposure
times given are per pixel in the resulting $512$-point spectrum. The
final column gives the $3\sigma$ upper limits to narrow line fluxes in
the spectra.  Note that due to poor atmospheric transmission and high
thermal background, the line flux limits do not hold in the regions
1.8 -- 2.1 $\mu$m and $> 2.35 \mu$m, and do not hold in regions
co-incident with the strongest sky emission lines.}}

\normalsize
\end{table*}

We observed 5 of the 7 galaxies without optical spectroscopic
redshifts with the CGS4 near-infrared spectrograph on UKIRT. The aim
of these observations was to search for emission lines or continuum
breaks to determine the redshifts. All the targets were observed in
the $H$-band and some were also observed in the $K$-band. $H$-band was
chosen because it is where one would expect to observe H$\alpha$ over
much of the redshift range with no bright lines in the optical ($1.3
\leq z \leq 1.8$). Details of the observations are given in Table
\ref{tab:cgs4}. The observing method used was that of Eales \&
Rawlings (1993) and involved 2x2 sampling, which achieves Nyquist
sampling by stepping the array by 0.5 resolution elements and ensures
that each wavelength is sampled by 2 pixels (to facilitate bad pixel
removal). Conditions were photometric throughout.

The observations were reduced in a standard way. Briefly, the steps
involved were (i) combine flat-fielded observations taken at different
positions along the slit; (ii) wavelength calibrate using sky OH
emission lines and subtract residual background; (iii) extract 2 pixel
(1.2 arcsec) and 6 pixel (3.6 arcsec) wide aperture spectra from the
2-D images; (iv) combine `positive' and `negative' spectra; (v)
flux-calibrate using standard stars; (vi) remove atmospheric
absorption by dividing by normalised A or F stars.

Weak continua were detected in all these spectra. No obvious breaks in
the continua were found, but our limits on any break are not very
stringent. For two objects, possible emission lines were observed in
the $H$-band spectra. For the other objects we are able to place
limits on any emission lines in regions covered by the spectra. For
5C6.62, the possible line is at $1.612 \pm 0.002 \mu$m and has a FWHM
velocity width of 1700 km s$^{-1}$. The line flux is $5.4 \times
10^{-19}$ W m$^{-2}$, so the detection is at the 3$\sigma$ level. For
5C7.245, the observed probable line is also at about the 3$\sigma$
level with a flux of $4.5 \times 10^{-19}$ W m$^{-2}$. The line centre
is $1.712 \pm 0.002 \mu$m and the FWHM is 1200 km s$^{-1}$. In both
cases the line widths are at roughly the level of the instrumental
resolution, so they may be intrinsically narrower than this. Fig.
\ref{fig:specgs4} shows the extracted spectra with these features
marked. The most likely emission line to detect in $H$-band is
H$\alpha$, because of the suspected redshifts of $z > 1$ for these
objects. Making the assumptions that these lines are real and
identifying them with H$\alpha$ leads to redshifts of $z=1.456 \pm
0.003$ for 5C6.62 and $z=1.609 \pm 0.003$ for 5C7.245. In Section
\ref{notes} we will discuss how these values fit with other evidence.

The expected H$\alpha$ fluxes for 7CRS radio galaxies at $z\approx 1.5$
can be determined from the emission line --- radio correlation
discussed in Willott et al. (1999). It is found that the typical flux
should be $\sim 3 \times 10^{-19}$ W m$^{-2}$. This is similar to the
limits obtained from spectroscopy. Therefore our inability to
detect lines in most sources does not rule out that they have
H$\alpha$ in the $H$-band. In fact, this shows that our detection of a
line in two sources is just as expected, given the scatter in this
relationship. More importantly, the lack of very strong lines
dominating the near-IR emission (and the spectroscopic detection of
continua in all cases) enables us to be confident about modelling the
SEDs of these sources in terms of continua.

\begin{figure}
\hspace{-0.25cm} \epsfxsize=0.48\textwidth \epsfbox{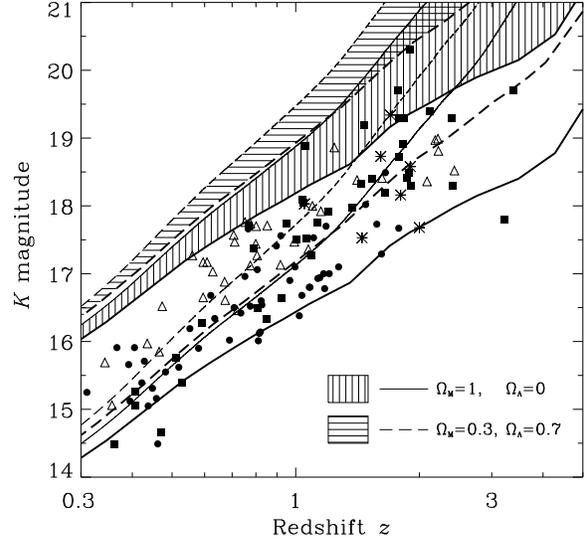}
{\caption[junk]{\label{fig:kz} $K$-band Hubble diagram for radio
galaxies in the 3CRR (filled circles), 6CE (filled squares; Eales et
al. 1997 and Rawlings et al. 2000) and 7CRS (open triangles; Willott
et al., in prep.) complete samples (quasars and broad-lined radio
galaxies are not plotted). The seven 7CRS radio galaxies with
redshifts estimated in this paper are shown as asterisks. The thin
curves show non-evolving (but $k$-corrected), old (13 Gyr) galaxies
with luminosity $L^*$ (upper curves; assuming $M^*_K=-24.6$ from
Gardner et al. 1997) and $5L^*$ (lower curves) for $\Omega_{\mathrm M}
=1$, $\Omega_ \Lambda=0$ (solid) and $\Omega_{\mathrm M}=0.3$,
$\Omega_\Lambda=0.7$ (dashed). Models accounting for passive evolution
are also shown for each luminosity and cosmological model (thick
lines) assuming all the stars in the galaxies formed in an
instantaneous burst at $z=10$. For the $L^*$ models we also show the
shaded region between the no-evolution and passive evolution models
for each cosmology. }}
\end{figure}

\section{SED fitting} 
\label{sedfitting}

The tightness of the near-infrared Hubble diagram (also known as the
$K-z$ relation) for 3C radio galaxies (Lilly \& Longair 1984) has led
to its widespread use as a redshift indicator for radio galaxies for
which no spectroscopic redshift has been obtained (e.g. Dunlop \&
Peacock 1990). However, Eales et al. (1993) found that there is
increased scatter beyond $z=2$ and this limits its usefulness at faint
magnitudes as a redshift indicator. In Fig. \ref{fig:kz} we show the
$K-z$ relation for the 3CRR, 6CE and 7CRS complete samples of radio
galaxies. Eales et al. (1997) found that 6CE galaxies at $z\ge 0.6$
were systematically fainter at $K$-band than 3C galaxies of the same
redshift. This correlation between $K$-band luminosity and radio
luminosity is confirmed with the addition of the 7CRS data
(Fig. \ref{fig:kz}) and appears to be present across the whole
redshift range of the sample. This casts serious doubt on naive use of
the $K-z$ relation as a redshift indicator.

Observations show that the $K$-band light is dominated by stars for
most of the objects on this plot, with negligible non-stellar
contamination in most cases (e.g. Best et al. 1998; Simpson et
al. 1999). Since the $K$-band light of galaxies is dominated by stars
of approximately solar mass (which make up the bulk of the total
stellar mass of galaxies), the $K$-band luminosity is a good tracer of
the total stellar mass of the galaxies. The model curves on
Fig. \ref{fig:kz} show non-evolving and passive evolution $L^*$ and
$5L^*$ galaxies as a function of redshift. The passive evolution model
assumes the stars in the galaxies formed at $z=10$.

It is clear that there is some evolution in the $K-z$ relation for
radio galaxies, especially for low values of $\Omega_{\rm
M}$. However, as noted by Lilly \& Longair (1984) and Eales et
al. (1997), passive evolution of galaxies formed at high redshift ($z
\gtsimeq 5$) is sufficient to explain the observations. We find that a
range of luminosities between $L^*$ and $5L^*$ is required to account
for the radio galaxies. The high luminosities show that all these
radio galaxies are very massive ($\gtsimeq 10^{11} {\rm M}_\odot$ in
stars) --- a feature which is key to our discussion in Section 7. The
$K-z$ diagram for the 3CRR, 6CE and 7CRS samples will be discussed in
more detail in a future paper (see also Lacy, Bunker \& Ridgway 2000).

Given the fairly large scatter in this relationship at high-redshifts
we conclude that the only reliable redshift constraints that can be
inferred are $z<1.3$ for $K<17$ and $z>1$ for $K>18$. It is clear that
for objects fainter than $K=17.5$ (i.e. all those in this study) no
firm upper limit on the redshift can be obtained from the $K-z$
relation (particularly because at the highest redshifts, the $K$-band
is actually probing optical light which is far more easily
contaminated by non-stellar light). It is for this reason that we have
obtained the data presented here in order to better constrain the
redshifts of these radio galaxies.

The imaging data allow us to place strong constraints on the
observed-frame optical--near-IR spectral energy distributions of these
extremely red objects. If one assumes that the observed SED is
dominated by light from a stellar population in the galaxy, then the
redshifts can be well-constrained (possible deviations from this
assumption will be discussed in Section \ref{notes}). Often just three
variables are considered in this type of work: redshift, age of the
stellar population and a normalisation (equivalent to the luminosity
or mass of the galaxy). However, in the light of the continued debate
about the nature of EROs, it is also beneficial to include an extra
variable --- the reddening of the SED by dust.

In the fitting procedure we describe here, a sample of 12 simple
stellar populations (instantaneous burst) from the GISSEL96 models of
Bruzual \& Charlot (in prep.) are adopted for the galaxy SEDs. These
models have ages of 0.01, 0.05, 0.1, 0.3, 0.6, 1.0, 1.4, 2.0, 2.5,
3.0, 3.5, 4.0 Gyr. The range of ages adopted were chosen because of
the expectation that many of the galaxies would be old ellipticals.
Therefore we want increased time resolution (in terms of the logarithm
of the age) at large ages, especially since the shape of the SEDs does
not change greatly for ages above 2 Gyr. We only consider solar
metallicities, since there is a well-known degeneracy between
metallicity and age; one should always bear in mind that an old galaxy
SED can probably be equally well-fit by a younger galaxy with higher
metallicity (Worthey 1994). We have not incorporated more complex
models such as ongoing star-formation or combined multiple-age
populations, since this would lead to a large increase in the number
of models. For our purposes the instantaneous burst models should be
sufficient, since they give a lower limit to the age.

The question of what effect dust-extinction has on the observed SED of
a dusty galaxy is complex and poorly understood. Scattering and
absorption, combined with the uncertainty of the dust distribution
relative to the light distribution as a function of wavelength,
complicate this issue. We assume here that the effects of dust
extinction follow the extinction law of Calzetti (1997). We consider
extinctions in the range $0 \leq E(B-V) \leq 1$ in steps of 0.1; for
the Calzetti law, the visual extinction $A_V=4E(B-V)$.

For the above range of models, we have calculated the expected
observed $R-K$, $I-K$, $J-K$ and $H-K$ colours as a function of
redshift, taking into account the filter/detector response. Thus for a
range of model $K$-magnitudes (within $\pm 2 \sigma$ of the observed
$K$-magnitude), we have model magnitudes in all bands. These were then
converted to fluxes and compared to the observed fluxes using a
$\chi^2$ test to determine the goodness-of-fit as a function of
redshift, age and reddening. In the case of non-detection of the
object in one of the filters, the flux in this filter was set to zero
and the photometric error obtained from the image. The reduced
$\chi^2$ was calculated by dividing by the number of data points.

\begin{figure*}
\epsfxsize=0.95\textwidth \epsfbox{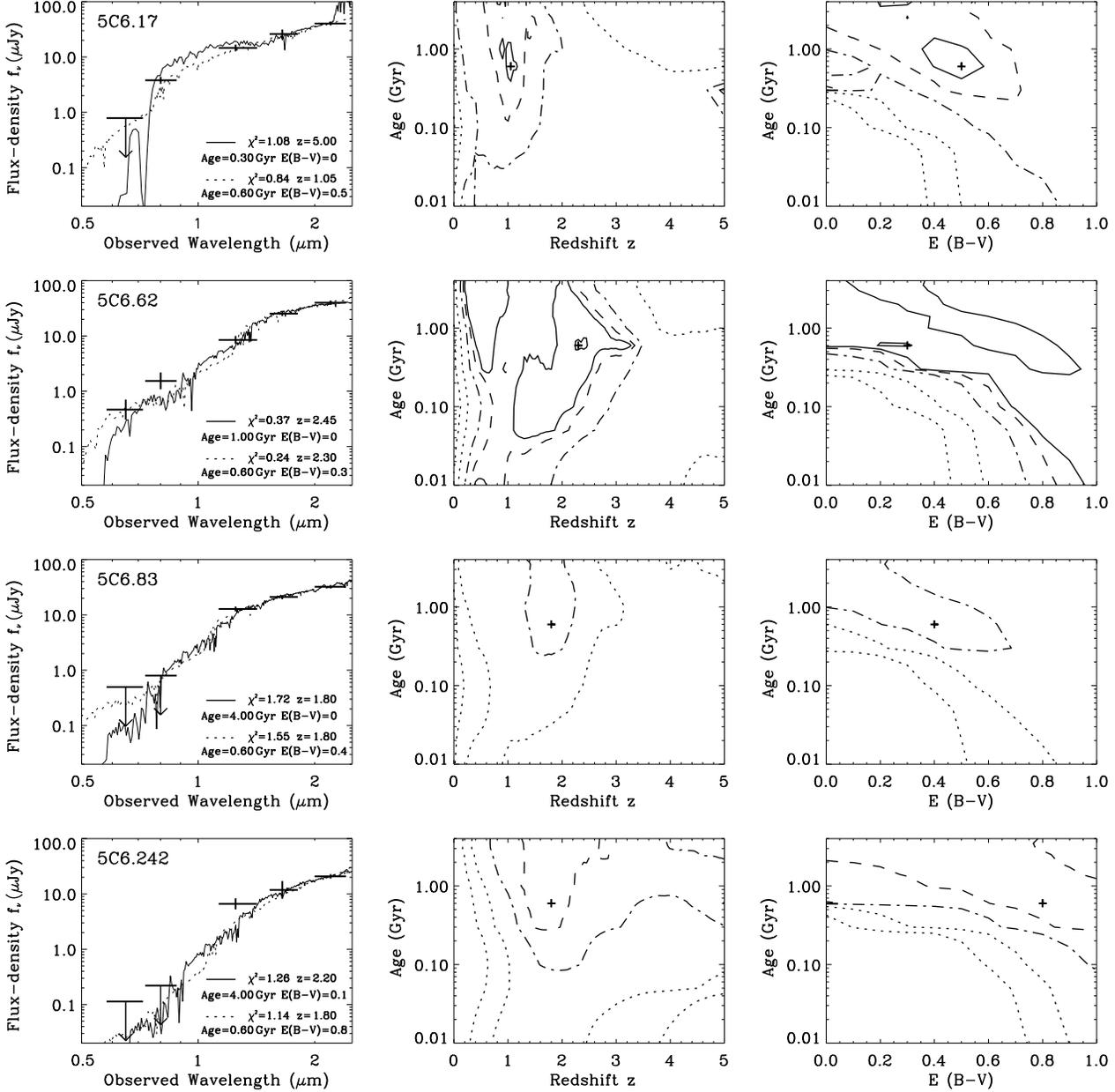}
{\caption[junk]{\label{fig:sed5c6} Results of the model-fitting to the
optical to near-IR SEDs of 7C-I (5C6) radio galaxies. The left panel
shows the broad-band photometric data with best-fit model
galaxies. The best fit unreddened [defined as $E(B-V)\leq 0.1$] model
is shown as a solid line and the best-fit reddened model as a dotted
line. Details of these fits are given in the bottom-right corner. The
centre and right panels show how the $\chi^2$ of the fit depends upon
redshift, age of the stellar population and reddening. The line style
used indicates the value of the reduced $\chi^2$: solid lines 0.3, 1.0
(where appropriate); dashed 1.5; dot-dashed 3.0; dotted 10, 30. The
peak in reduced $\chi^2$ is shown as a cross.}}
\end{figure*}

\begin{figure*}
\epsfxsize=0.95\textwidth 
\epsfbox{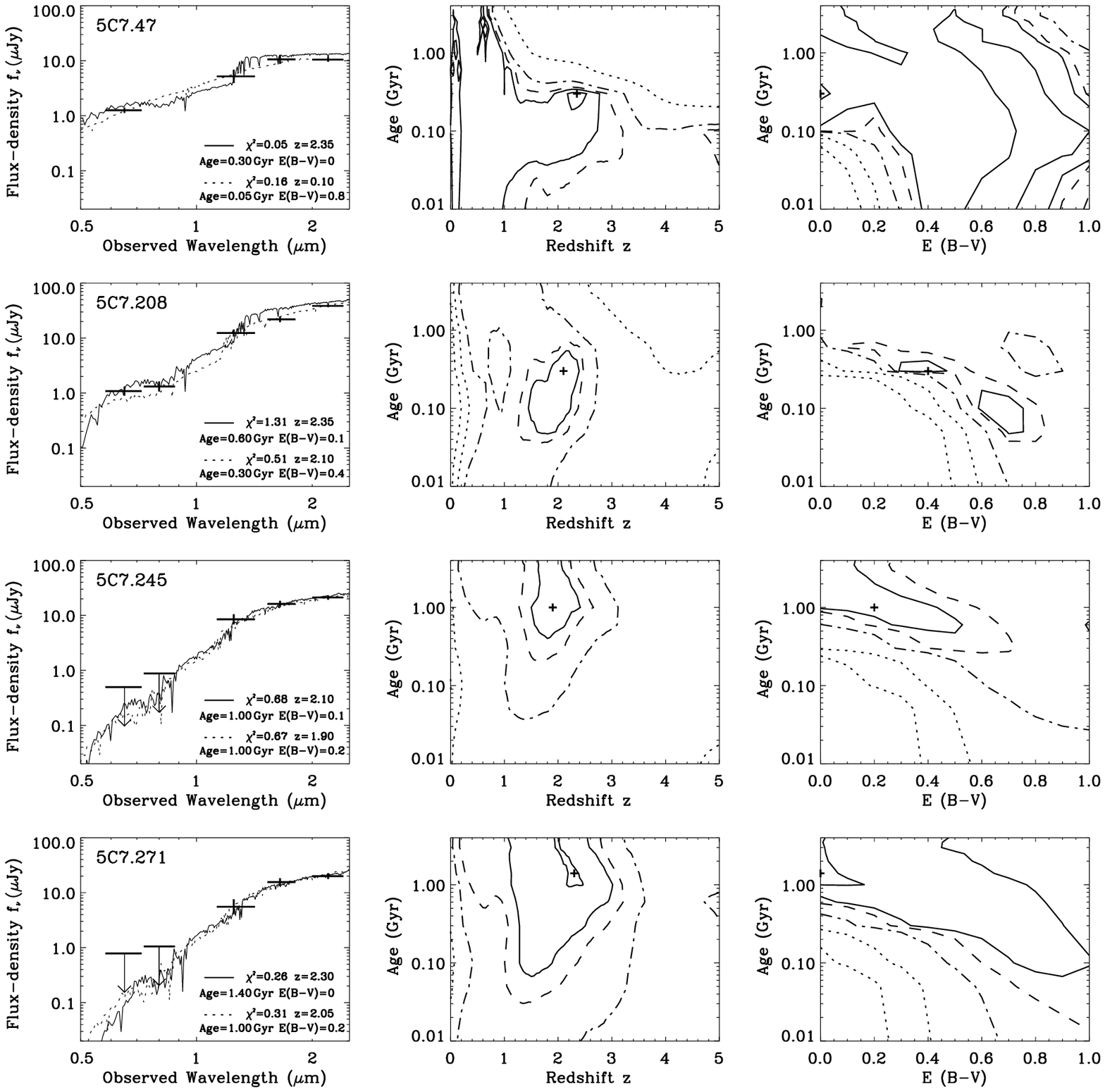}
{\caption[junk]{\label{fig:sed5c7} As Fig. \ref{fig:sed5c6} but for
the 7C-II (5C7) galaxies.}}
\end{figure*}

Figs. \ref{fig:sed5c6} and \ref{fig:sed5c7} show the results of this
fitting process for the 7C---I (5C6) and 7C---II (5C7) sources
respectively. For each source the left panel shows the optical and
near-IR photometry, overlaid with the best-fit reddened [$E(B-V)
\geq 0.2$ -- dotted line] and unreddened [$E(B-V) \leq 0.1$ -- solid
line] galaxy models. The parameters of the best-fits are given in the
bottom-right corner of the panel along with the reduced $\chi^2$. The
centre and right panels show the reduced $\chi^2$ contours as a
function of age and redshift (centre) and age and reddening
(right). Fits with reduced $\chi^2 < 1$ were found for six objects and
the remaining two had best-fit reduced $\chi^2 < 2$. To estimate the
errors on the best-fit parameters we consider all models with reduced
$\chi^2 \leq 1.5$ as acceptable.

The results for each object will be discussed in the following
section. A general point to note is the fairly wide range of
parameters that provide acceptable fits for the objects. The
degeneracy of age and reddening appears to be the strongest cause of
this as shown by the diagonal shape of contours in the age-reddening
plane (from old and dust-free at top-left to young and dusty at
bottom-right). In contrast there is little correlation between
redshift and age in the contours and the best-fit redshifts are in
most cases similar for both the reddened and unreddened cases. Thus we
will be able to provide strong constraints on the redshifts of these
sources. However, the degeneracy between age and reddening (and also
that between metallicity and age) make it very difficult to firmly
constrain the ages of these objects from broad-band photometry alone.

\section{Notes on individual sources} 
\label{notes} 

{\bf 5C6.17} 

This is a large (47 arcsec) double radio source with a probable but
weak radio core (detected at the 3$\sigma$ level). The presence of
extended lobe emission and the large angular size strongly suggest a
redshift $z \ltsimeq 2$ (Blundell, Rawlings \& Willott 1999).  The
near-IR counterpart is at the core position (within the uncertainties
of the registration of the radio and optical astrometry frames,
$\approx 1$ arcsec). This object is clearly extended in the 1 arcsec
seeing $J$ and $H$ images. It lies in a fairly empty field with no
companion galaxies detected in any of the images within a 10 arcsec
radius.

The non-detection at $R$-band gives $R-K>6.0$ (5 arcsec diameter
aperture magnitudes are used here because the source is quite
extended). However, its detection at $I$-band gives $R-I>1.8$ which
indicates a steepening of the SED between $I$ and $R$. This steepening
is best interpreted as the 4000 \AA~ break falling between these
bands. The SED fitting procedure supports this conclusion with a
best-fit redshift of $z=1.05$. The SED is fit slightly better by a
heavily reddened stellar population and a 0.6 Gyr galaxy reddened by
$E(B-V)=0.5$ provides the best-fit, but no reddening and an age of 4
Gyr is still acceptable. There is a secondary minimum in the $\chi^2$
contours at $z=5$, but we think this redshift extremely unlikely given
the radio structure, the $K$ magnitude and the lack of a Lyman-$\alpha$
emission line in the optical spectrum. With a 5 arcsec $K$ magnitude
of $18.0$, 5C6.17 falls at the faint edge of the $K-z$ relation at
$z=1$. For redshifts $z<0.6$, this source would be over a magnitude
fainter than the mean relation, so such low redshifts are unlikely on
these grounds as well as being ruled out by the SED fitting.

At $z\approx 1$ we would expect to have detected [OII] from the narrow
line region in our optical spectrum with a flux in the range $1 - 5
\times 10^{-19} {\rm W \, m}^{-2}$ (Willott et al. 1999). The 3-$\sigma$
limit on any narrow emission line in our 1 hour WHT spectrum is $4
\times 10^{-20} {\rm W \, m}^{-2}$. Therefore reddening of the narrow
line emission by $E(B-V) > 0.2 - 0.5$ (depending upon the intrinsic
line luminosity) would be sufficient to render the line undetectable
in our spectrum.  In the case of no reddening, then the emission line
strength of this object would have to be particularly weak relative to
the other objects in the 7CRS at $z=1$.  It should also
be noted that a redshift in the range 1.04 to 1.06 would place the
[OII] line in the region of high atmospheric absorption and slightly
higher redshifts possibly into the strong OH sky lines. Hence our
inability to detect the [OII] emission line if 5C6.17 is at
$z\approx1$ is not a serious concern.

5C6.17 has a redshift constrained to $0.6 \leq z \leq 1.2$ and the
best-fit value of $z=1.05$ is adopted.

\vspace{0.3cm}

\noindent {\bf 5C6.62} 

This is a double radio source with a weak core and angular size of 34
arcsec (Blundell et al., in prep.). The position of the
near-IR/optical identification is coincident with the radio
core. There is significant extended emission around the hotspots, one
of which contains no compact features. These radio properties are
suggestive of $z \ltsimeq 2$. The radio source identification is
clearly extended in the 1 arcsec seeing $H$ and $K$ images and there
are hints in the $J$ and $K$ images of an extension/companion to the
south-east, although better images would be necessary to confirm this.

The galaxy is detected in all the filters, although the $R$ and
$I$-band detections have low signal-to-noise ratios (SNR); 3 arcsec
aperture magnitudes are used to improve the SNR. 5C6.62 has a very red
$R-K$ colour of 6.6. Between the $R$ and $H$ bands the SED is
virtually a power-law with a slight flattening of the slope towards
$K$. The power-law slope with no sign of a break indicates that
dust-reddening may be important in this object. Due to the rather
large errors on the optical data, there are a wide range of models
which provide acceptable fits and the $\chi^2$ plots are complex.  The
best-fit models are at $z=2.4$, but the formal redshift range allowed
by the fitting is $0.2 \leq z \leq 3.3$. Only higher redshifts are
strongly ruled out.

This is one of the two sources to have probable emission lines in
their near-IR spectra. Taking the line to be H$\alpha$ gives a
redshift of $z=1.45$. Note that 5C6.62 is the brightest object at
$K$-band out of those studied in this paper, with $K=17.5$. The fact
that it is resolved removes the possibility that a quasar component
dominates. On the $K-z$ relation, this magnitude corresponds to
7CRS/6CE sources with $0.7 \leq z \leq 1.3$. The evidence from the
$K-z$ relation would appear to rule out the higher redshifts ($z
\gtsimeq 2$) compatible with the fitting. Assuming the redshift of
$1.45$ from the spectroscopy, Fig. \ref{fig:sed5c6} shows the highest
likelihood at this redshift ($\chi^2 \leq 1$) is for a young (100 Myr)
galaxy with very high reddening [$E(B-V) \approx 0.9$]. This is
clearly ruled out on luminosity grounds since it implies 2 magnitudes
of reddening at the observed $K$-band and an intrinsic $K=15.5$. This
is clearly not feasible for a $z=1.5$ radio galaxy. However,
Fig. \ref{fig:sed5c6} also shows that a wide range of ages at $z=1.45$
fall within the $\chi^2 \leq 1.5$ region and therefore zero (or at
least minimal) reddening is also acceptable.

The redshift of 5C6.62 is not well-constrained by the SED fitting. We
adopt a redshift of $z=1.45$ based on the possible emission line
detection in the $H$-band spectrum. Taking into account the $K$-band
luminosity we find that the possible redshift range is $0.5 \leq z
\leq 2.0$.

\vspace{0.3cm}

\noindent {\bf 5C6.83} 

This is a 14 arcsec double radio source (Blundell et al., in
prep.). No radio core is detected, but the assumed near-IR counterpart
lies midway between the two hotspots. The radio source lies in a
crowded field with many objects visible in the optical and near-IR
images. The brightest object is a star, but the next two brightest
objects are both extended and would appear to be low redshift galaxies
in a group/cluster. It is not possible to tell if all the fainter
objects are resolved or not, but it appears that the radio source
counterpart is slightly extended in the near-IR images.

Because of the good seeing in all images and fairly compact nature of
the galaxy, 3 arcsec aperture photometry is used. The galaxy is
undetected in the $R$ and $I$ images and has a colour $R-K>6.2$. From
the $J$ to $K$ bands the SED is equivalent to that of a galaxy
population beyond the 4000 \AA~ break. This constrains the redshift to
$z<2.2$. The sharp drop down to the faint $I$ and $R$ limits, combined
with the near-IR data, can be fit by either an old stellar population
with no dust reddening or a younger, heavily reddened population.
Note that the best-fitting model has a reduced $\chi^2$ of 1.55, just
beyond that formally acceptable.  This model has an age of 0.6 Gyr and
$E(B-V)=0.4$. The degeneracy between age and dust allows older,
less-dusty models with only a small increase in $\chi^2$. The redshift
for this galaxy is well-constrained by the fitting process and
although no formal range can be determined, Fig. \ref{fig:sed5c6}
suggests that a range of $1.5 \leq z \leq 2.1$ is appropriate.

The $K$ magnitude of 18.0 is consistent with $z \approx 1.8$ according
to the $K-z$ relation.  This redshift is in the difficult region
without bright emission lines in optical spectra. The fitting allows
redshifts up to $z=2.1$. Redshifts in the range $1.8 < z \leq 2.1$ may
be possible, but they would require a Ly$\alpha$ line which is either
intrinsically weak or extinguished by dust (e.g. Charlot \& Fall
1993). We are left with the redshift range of $1.5 \leq z \leq 2.1$,
with a most-likely value of $z=1.8$.

\vspace{0.3cm}

\noindent {\bf 5C6.242} 

This is a small (5 arcsec) double radio source with a weak core. The
near-IR counterpart is co-incident with the radio core. The Keck-II
$K$-band image with 0.6 arcsec resolution shows the counterpart to be
resolved but fairly compact. There is another extremely red galaxy 7
arcsec south-west of the radio source. The bright object just north of
the radio source is a star.

5C6.242 is not detected in our deep WHT $R$-band image and there is
only a hint of the object in the Keck $I$-band image which is not
statistically significant. It is detected in all the near-IR images.
3 arcsec aperture photometry is used because of the low
SNR of the $J$ and $H$ detections. This is the reddest of
all the objects studied in this paper with $R-K>7.3$, although
this may be because it has the deepest $R$-band data and several other
sources have only lower limits on their colours. The shape of the
photometry from $J$ to $K$ is similar to that of a galaxy above the
4000 \AA~ break, implying a redshift $z<2.2$. The flux-density then
plummets down to the $I$ and $R$ bands. The $I-J$ colour of $>4.3$ is
astonishing. The best-fit ($\chi^2 =1.1$) model has a redshift of
$z=1.8$, age $= 0.6$ Gyr and $E(B-V)=0.8$. As with 5C6.83, an
unreddened old galaxy gives a marginally worse fit. However, this
model has an age of 4.0 Gyr at a redshift of 2.2, which is not allowed
by some cosmological models (see Section 7.1).

With a $K$-magnitude of 18.5, the $K-z$ relation suggests that 5C6.242
is extremely unlikely to be at $z<1$. This is supported by the SED
fitting which also excludes $z<1$ models at a high level of
confidence. As with 5C6.83, we use Fig. \ref{fig:sed5c6} to get a
rough range of allowed redshifts from the fitting.  We conclude that
the redshift of 5C6.242 lies in the region $1.3 \leq z \leq 2.6$ with
a most-likely value of $z=1.9$ (although this requires
weak Ly$\alpha$).

\vspace{0.3cm}

\noindent {\bf 5C7.47} 

The optical/near-IR counterpart is co-incident with the position of
this compact, but resolved (angular size $\sim 0.2$ arcsec) radio
source (Blundell et al., in prep). This object is probably resolved in
the good-seeing $R$, $J$ and $H$-band images.  There is a bright
($R=19.8$) galaxy 7 arcsec west of the radio source which has optical
and near-IR colours consistent with a galaxy at $z=0.4 \pm 0.2$. There
are two very red ($R-K>5$) objects within 20 arcsec of 5C7.47. The
optical spectrum of one of these objects (ERO J081539+2446.8) has an
emission line which is likely to be [OII] at $z=1.200$. This object is
discussed in detail in the appendix. Within 30 arcsec of 5C7.47 there
are a large number of faint galaxies visible in both the $R$ and
$K$-band images. These galaxies have magnitudes of $K\approx 20$ and
$R \approx 24$. They would appear to be part of a cluster, although
with their bluer colours than ERO J081539+2446.8 they may not be at
the same redshift. In fact, they have similar colours to 5C7.47 and
the cluster may host the radio source.

5C7.47 is unusual in our sample studied here because it is both the
bluest galaxy with $R-K = 4.0$ and the faintest at $K$-band with a
magnitude of 19.5 (3 arcsec aperture photometry used). The optical
spectrum (60 minutes integration) shows a faint continuum all the way
down to 3900 \AA. Hence we can be sure of a redshift below 3.3 because
of the absence of any observed Lyman limit. There are no clear
emission lines in the spectrum to a 3 $\sigma$ limit of $2 \times
10^{-20} {\rm W m}^{-2}$. Note that we do not have any $I$-band data
for 5C7.47, so constraining the redshift is difficult. In addition the
relative brightness at $R$-band compared to that at $K$ may be
indicating that the $R$ flux is dominated by non-stellar emission from
the AGN. If this is the case, then the $R$-magnitude cannot help us
constrain the redshift and we are left with the $J,H,K$ data.

\begin{figure*}
\vspace{-7.6cm}
\epsfxsize=1.0\textwidth \epsfbox{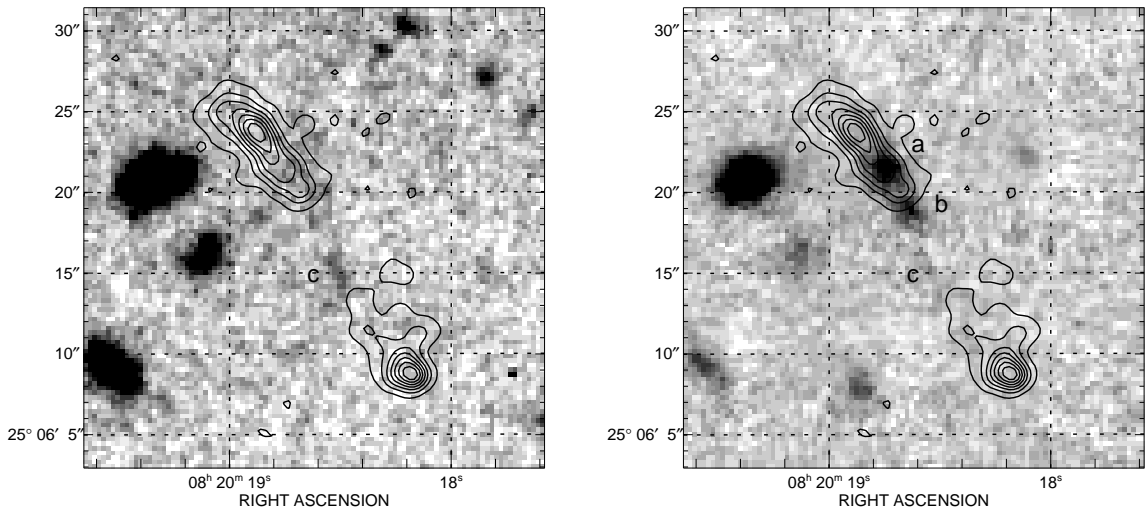}
\vspace{-8.2cm}
{\caption[junk]{\label{fig:7p208} Radio contours at 5 GHz overlaid on
greyscales of the $R$-band (left) and $K$-band (right) images of
5C7.208. Components `a', `b' and `c' described in the text are
labelled. `a' and `b' have $R-K>5$ so consequently are relatively
bright in the $K$ image and very faint in the $R$ image; component `c'
is much bluer. We suspect that either `a' or `b' is the true
counterpart to the radio source and their similar colours suggest they
are at the same redshift. }}
\end{figure*}

The $\chi^2$ contours from the SED fitting are rather complex and have
low values at a range of redshifts. The best-fit model is a young
dust-free galaxy at $z=2.35$. The $\chi^{2}$ rapidly increases at
redshifts greater than this, due to the fact that the $J$ flux-density
is not greatly down on that at $H$ and $K$. This would seem to be a
fairly firm limit (as it is for all the objects discussed in this
paper).  If one assumes that the $R$-band point is due to
contamination from an AGN (and adopt a limit of $R \ge 25$ for the
stellar population), the best-fit redshift would be $1.8 \pm 0.3$. The
lower redshift limit cannot be determined by the SED fitting because
of the small $\chi^{2}$ at all low redshifts. However, the fact that
5C7.47 is the faintest at $K$-band suggests a fairly high redshift and
the $K-z$ relation shows that $z<1.5$ is unlikely. Note that this
explicitly assumes the host to be a massive galaxy which at least
approximately follows the $K-z$ relation. If instead the host is a
much less luminous galaxy then it could be at any low redshift (but
the compact radio structure with little extended emission would be
very unusual for such a low-redshift/luminosity radio
galaxy). Adopting this assumption we find that the possible range of
redshifts for 5C7.47 is $1.5 \leq z \leq 2.5$. Once again weak
Ly$\alpha$ emission would be required for $1.8 \leq z \leq 2.5$, which
would be highly unusual for a powerful radio galaxy with an observed
continuum down to virtually the wavelength of redshifted
Ly$\alpha$. Therefore we assign a redshift of $z=1.7$ to 5C7.47.

\vspace{0.3cm}

\noindent {\bf 5C7.208} 

This is an 18 arcsec double radio source with some rather diffuse lobe
emission near both hotspots (Blundell et al., in prep.). It is the
only source discussed in this paper which shows clear evidence for a
radio-optical alignment. There are two near-IR components separated by
2.9 arcsec at a position angle (east of north) of $34^{\circ}$. This
is within a couple of degrees of the radio axis (Fig.
\ref{fig:7p208}).  We define the brighter (at $K$-band) northern
component `a' and the fainter southern component `b'. There is more
faint near-IR emission extending $\approx 5$ arcsec further south-west
along a similar position angle, which we refer to here as component
`c'. The colours of `a' and `b' are very similar and very
red. Component `c' has a much bluer colour and is brighter in the
optical than `a' or `b'. Lacy et al. (1999a) find some residual
alignment effect (often occurring in discrete blobs) in
low-luminosity radio sources such as those in the 7CRS.  There are
several bright, fairly blue galaxies in the field which would appear
to be a group at low redshift.

Component `a' is detected at reasonable signal-to-noise in all the
images -- 3 arcsec aperture photometry has been performed since the
seeing was similar in all bands. `b' is just visible in the $R$ and
$I$ images but not at a sufficient significance level, so upper limits
to these fluxes have been adopted. The optical spectra of Willott et
al. (in prep.) shows weak, patchy continuum from `c', but nothing at
all from `a' or `b'. No significant emission lines are observed in
these spectra. We only consider here the SED of component `a' since we
suspect it is the radio source host galaxy (the position in Section 2
and magnitudes in Table 1 refer to this object). Note that since the
colours of `a' and `b' are very similar, the modelling described is
also applicable to `b'. `a' has $R-K=5.6$ and is well-fit by a
reddened [$E(B-V)=0.4$] young (0.3 Gyr) galaxy at $z=2.1$. Unreddened,
older galaxies at slightly higher redshifts are also not ruled
out. The formal redshift range allowed is $1.3 \leq z \leq 2.45$.

The flatness between $R$ and $I$ could be indicating a non-stellar
component, but it is also fit by stellar populations as shown in
Fig. \ref{fig:sed5c7}. It is possible that this is a case of an
alignment effect, but the spectral shapes of both `a' and `b' are
strongly suggestive of stellar populations and not scattered quasar
light. The most likely answer is that `a' and `b' are a pair of
galaxies at the same redshift, which just happen to be aligned with
the radio axis (see West 1991 and Eales 1992 for possible explanations
of why companion galaxies may preferentially lie along the radio axes
of radio galaxies). The $K$-magnitude of `a' is 17.7, consistent with
a redshift $z > 0.6$. 

We conclude that the redshift of 5C7.208 falls somewhere in the range
$1.3 \leq z \leq 2.4$ and adopt $z=2.0$ from the SED fitting. Note
that this fit is for a dusty galaxy and this might explain the lack of
Ly$\alpha$ in our optical spectrum.

\vspace{0.3cm}

\noindent {\bf 5C7.245} 

This is another small (12 arcsec) double radio source (Blundell et
al., in prep.). The north-west hotspot has a strange structure with two
peaks in intensity separated by 3 arcsec. The near-IR identification
is very close to the southern hotspot, along the radio axis, and
definitely extended in the 0.8 arcsec seeing $J$ and $H$ images. The
radio source lies in a fairly blank field with few other galaxies
detected on our images.

Undetected in the $R$ and $I$ bands, 5C7.245 has $R-K>5.8$ (5 arcsec
apertures used because of the varied seeing of the observations).  As
for most of the objects previously discussed, the spectrum is fairly
flat between $J$ and $K$ and then drops rapidly to $I$ and $R$. The
SED fitting suggests an unreddened galaxy at $z \approx 2$ with a
formal range of $1.3 \leq z \leq 2.65$.

The near-IR spectrum shows a probable emission line which we believe
to be H$\alpha$ at $z=1.61$. The SED fitting and $K$-magnitude are
fully consistent with this.

\vspace{0.3cm}

\noindent {\bf 5C7.271} 

This is a very small but resolved ($\sim 1$ arcsec) radio source
(Blundell et al., in prep.). The near-IR counterpart is co-incident
with this radio position and is resolved in the 0.8 arcsec seeing $J$
and $H$ images. The few other objects in the field appear to be
relatively low-redshift galaxies.

5C7.271 is undetected in the $R$ and $I$ band images and has a lower
limit on its colour of $R-K>5.2$ (5 arcsec apertures used because of
the varied seeing of the observations). The SED fitting gives a wide
range of acceptable redshifts ($0.2 \leq z \leq 3.2$) with a
best-fit value of $z=2.3$. As mentioned in section \ref{sample}, we
acquired a spectroscopic redshift for this source of $z=2.224 \pm
0.006$ after we had obtained this imaging data. Since it has colours
similar to the other objects studied here, we have used it to check
the reliability of our SED fitting procedure. The similarity of the
SED-fit redshift to the spectroscopic one is encouraging. Note that
for this object there is quite a jump between $J$ and $H$, compared to
the other sources studied here. This is an indicator of its slightly
higher redshift since the 4000 \AA~ break has entered the $J$-band.

\begin{figure} 
\hspace{-0.25cm} 
\epsfxsize=0.48\textwidth
\epsfbox{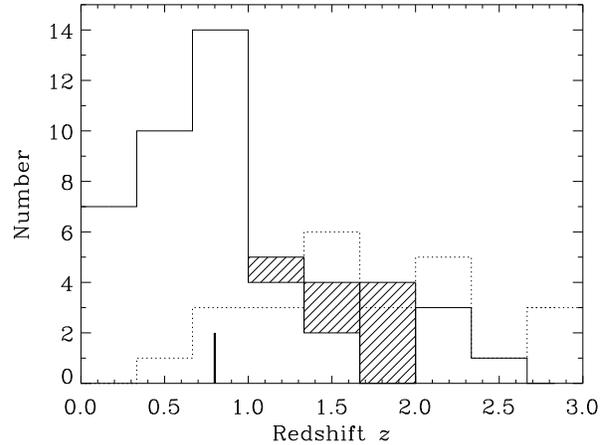}
{\caption[junk]{\label{fig:zdist} Redshift distribution of radio
sources in the 7C-I and 7C-II regions of the 7CRS. The solid line
gives the histogram for radio galaxies with spectroscopic redshifts
and the hatched areas the seven galaxies with redshifts estimated in
this paper. For comparison, the redshift distribution of quasars in
the sample is shown as a dotted histogram. The thick bar
shows the rough location of the redshift at which the radio galaxies
change from being of the low emission line luminosity class to the
high luminosity class as defined by Willott et al. (2000a).}}
\end{figure}

\section{Robustness of redshift estimates}

Constraining the redshifts of all the radio sources in complete
samples is extremely important. For example, only a few sources at
$z>3$ in a sample such as this would refute the tentative evidence for
a high-redshift decrease in the comoving space density of radio
sources (Dunlop \& Peacock 1990; Willott et al. 2000b). In Fig.
\ref{fig:zdist} we show the redshift distribution of radio galaxies
with spectroscopic redshifts in the 7C-I and 7C-II regions of the
7CRS. The seven sources with redshift estimates from this paper are
also shown. Note that the distribution for spectroscopic redshifts is
not smooth, with a peak at $z=0.8$ and another small peak at
$z=2.2$. Whilst redshift distributions such as this are not ruled out
by models of the radio luminosity function (for example the
dual-population model of Willott et al. 2000b predicts a peak at $z
\approx 0.8$ and another at $z \approx 1.6$ for a $S_{151} \geq 0.5$
Jy sample), the fact that the `hole' in the redshift distribution
occurs at $z \approx 1.5$ is suggestive of selection effects. This is
precisely the redshift range where there are no bright emission lines
in optical spectra. Hence it does not seem too surprising that our
estimated redshifts for the objects without narrow lines almost all
fall in this region. A similar gap is found in the redshift
distribution of sources in the 7C-III region, which has a similar
fraction of faint red galaxies without spectroscopic redshifts (Lacy
et al. 1999b, 2000).

It is informative to compare the redshift distribution of quasars in
the 7CRS, which should have a similar redshift distribution to the
radio galaxies according to unified schemes (at least for radio
luminosities $\log_{10} L_{151} > 26.5~ {\rm W Hz}^{-1} {\rm sr}^{-1}$
and hence $z>1$ in the 7CRS; see e.g. Willott et al. 2000a). For the
quasars (7C-I and 7C-II only), we find no such gap at these redshifts
(Fig. \ref{fig:zdist}). This is easily explained because the quasars
have strong broad C IV, C III] and Mg II lines in this redshift range,
making redshift determinations easy.

This analysis of the redshift distribution supports our redshift
determinations in Section \ref{notes} being predominantly in the range
$1< z <2$. Although for any individual object, there is a fair amount
of uncertainty (typically $\Delta z = \pm 0.3$), for the ensemble of
objects their redshifts are reasonably well-constrained. None of the
galaxies could plausibly be at $z>2.5$. At these high redshifts, all
of the $J$-band is below the 4000 \AA~ break and a large drop in
flux-density from $J$ to $H$ would be expected. The relative flatness
of the SEDs from $J$ through to $K$ (combined with the sharp decrease
in the optical) in all cases provides a very strong argument against
such high redshifts. Redshifts less than 1 are unlikely given the lack
of [OII] emission and the faint $K$-magnitudes of the galaxies.
Therefore we are confident that these redshift estimates are
reasonable and this gives the 7C-I and 7C-II samples virtually
complete redshifts, making them a powerful database for radio source
studies. Initial results using these datasets can be found in Blundell
et al. (1999) and Willott et al. (1998,1999,2000a,b).

\section{Discussion}

\subsection{Old or dusty high-redshift galaxies}

As we have seen in the fitting process, it is very difficult to be
confident about the ages of the galaxies, because of the
age--reddening degeneracy (and further uncertainty is due to the
age-metallicity degeneracy).  Spectroscopy of stellar absorption
features will be required to provide strong constraints on the ages
(e.g. Spinrad et al. 1997; but see below). This will be difficult even
with 8-10 m class telescopes because of the faintness of the galaxies
in the optical and the high sky background in the near-infrared. A
less robust but more feasible method would be measuring the position
of the 4000 \AA~ break which would give a very reliable redshift and
an indication of whether heavy dust-reddening is important
(e.g. Soifer et al. 1999). Sub-mm detections would be expected for
very dusty starbursts.

Two extremely red $z \approx 1.5$ radio galaxies from the LBDS survey
(LBDS 53W069 and LBDS 53W091) have Keck spectroscopy of their
rest-frame UV spectra (Spinrad et al. 1997; Dunlop 1999). Fitting
these spectra with stellar synthesis models gives ages of $\gtsimeq
3.5$ Gyr for both galaxies. The ages determined from fitting to the
spectral features (which are reddening--independent) are older than
those derived from continuum fitting -- indicating that for these
objects at least, reddening is not significant. Such a high age at
this redshift indicates a high formation redshift of $z_{\rm f} > 4$
(in both of the cosmologies used in this paper) and apparently rules
out cosmologies with simultaneously high values of both $H_0$ and
$\Omega_{\mathrm M}$. There has since been much controversy about the
true ages of these galaxies with more than one group showing that
alternative population synthesis models yield much smaller ages in the
range $1 - 2$ Gyr (e.g. Bruzual \& Magris 1997; Yi et al. 2000 -- but
see Nolan, Dunlop \& Jimenez 2000).  Therefore whilst it may be
premature to use such objects as reliable cosmological parameter
constraints, the existence of evolved ellipticals at these redshifts
do indicate high formation redshifts for most of their stellar
populations. Note that we have used the Bruzual \& Charlot models in
our analysis, which are at the lower end of the controversial range in
ages. Also, Spinrad et al. (1997) found that fitting to spectral
features in the rest-frame UV gives older ages than application of the
Bruzual \& Charlot models to broad-band colours. Therefore, our
derived ages are certainly lower limits.

Finding similar objects to the LBDS radio galaxies at higher redshifts
of $z\approx 2$ (such as several of the objects presented in this
paper) requires high formation redshifts even if these galaxies are 2
Gyr old, due to the rapidly decreasing time available at high
redshifts. For example, at $z=2$ an age of 2 Gyr corresponds to
$z_{\rm f} \gtsimeq 8$ for $\Omega_{\mathrm M}=1$, $\Omega_ \Lambda=0$
or $z_{\rm f} =3.5$ for $\Omega_{\mathrm M}=0.3$, $\Omega_
\Lambda=0.7$ (all assuming $H_0= 50$ km s$^{-1}$ Mpc$^{-1}$).  Note
that for a $\Omega_{\mathrm M}=0.3$, $\Omega_ \Lambda=0.7$ Universe
with an age of 13 Gyr (equivalent to that of the $\Omega_{\mathrm
M}=1$, $\Omega_ \Lambda=0$ Universe with $H_0= 50$) $H_0= 70$ km
s$^{-1}$ Mpc$^{-1}$ is required giving $z_{\rm f} =5$ .

Fig \ref{fig:zdist} shows that in the redshift range $1.3 \leq z \leq
2.0$, only two of the eight galaxies in the 7CRS had redshifts
determined from optical spectroscopy. These two galaxies, 5C6.217 and
5C7.57, do not have such red colours as most of the objects studied
here ($3 < R-K < 4$, Fig. \ref{fig:rkz}). The presence of
high-ionization lines in their optical spectra, such as C IV and [Ne
V], may suggest this blue light is non-stellar in origin. Together
with 5C7.47, we find that 3 of the 8 sources have fairly blue colours,
but the remaining 5 all have $R-K>5.5$. This shows that the UV
continuum emission in higher power radio galaxies, such as those in
the 3C sample, is negligible for the majority of 7CRS sources. At
$z\approx1$, Best et al. (1998) find the SEDs of 3C radio galaxies to
be consistent with old ellipticals once the aligned emission has been
accounted for. With just a modest increase in redshift, it would be a
surprise to find that they are suddenly all very dusty
starbursts. Sub-mm observations of 3C radio galaxies do not indicate a
change from old ellipticals to dusty starbursts over this redshift
range (Archibald et al. 2000).

Given the evidence above, we will hereafter make the assumption that
these red colours are due to old stellar populations and not
dust-reddening. Then we find that more than half of the galaxies
hosting radio sources at these redshifts must have formed at very high
redshifts (and if the three other objects are blue because of
non-stellar continuum, they too may have underlying old ellipticals).
These objects would appear to be very similar to low redshift
ellipticals with little current star-formation. Their formation
redshifts will be at $z_{\rm f} \gtsimeq 5$ for stellar ages $\gtsimeq
3$ Gyr (for $\Omega_{\mathrm M}=1$, $\Omega_ \Lambda=0$). There is no
evidence that these radio sources are in rich clusters from our
images, with just a couple of cases of possible companions at the same
redshift (5C7.208 and 5C7.47) but since we are only sensitive to
luminous galaxies at these redshifts, one cannot put strong limits on
the clustering around these sources.

The question then posed is if, as is normally assumed (e.g. Genzel,
Lutz \& Tacconi 1998), galaxy-galaxy mergers are the trigger for
powerful radio jets, why is there no sign of any associated
star-formation in these evolved galaxies?  Whilst the theory that most
high-redshift ($z > 0.6$) radio galaxies have recent or ongoing
massive star-formation (e.g. Chambers \& Charlot 1990) has been
challenged by the discovery that the UV-excesses in powerful radio
galaxies are, at $z \sim 1$, typically due to non-stellar mechanisms
(e.g. Cimatti et al 1997), there are certain clear-cut examples of
powerful radio galaxies with huge current star-formation rates,
e.g. the $z=3.8$ radio galaxy 4C41.17 (Dunlop et al. 1994; Dey et
al. 1997). A clue to understanding why intense star formation only
sometimes accompanies powerful radio jets may come from the very high
detection rates achieved by sub-millimetre photometry of the
highest-redshift radio galaxies (Archibald et al. 2000), suggesting an
especially strong association between powerful radio jets and intense
star formation in these systems.  These systems differ from, for
example the 7CRS galaxies studied here, in three crucial ways: (i)
they are observed at earlier cosmic epochs, when gas reservoirs in
galaxies were presumably less depleted (e.g. Storrie-Lombardi, McMahon
\& Irwin 1996); (ii) they possess intrinsically much more powerful
jets (e.g. Rawlings \& Saunders 1991), and are therefore plausibly
much more massive systems (e.g. Efstathiou \& Rees 1988); and (iii)
their radio sources are necessarily observed $\ll 10^7$ years after
the jet-triggering event (Blundell \& Rawlings 1999).  Some
combination of these three effects can easily explain the lack of
star-formation activity in the 7CRS EROs. Note that the radio source
ages for all the $R-K>5.5$ 7CRS sources studied in this paper lie in
the $10^7-10^8$ yr range (from equation 8 of Willott et al. 1999),
which are plausibly times long enough after the jet-triggering event
that any synchronized star formation has now faded (see also Best et
al. 1998).  Note that these radio source ages are in any case a factor
of more than 10 times less than the time since the cessation of the
{\em major} epoch of star-formation in these galaxies.

\subsection{Relationship to the LBDS EROs}

We now investigate how our finding of a large population of very red
galaxies at $1<z<2$ compares with the similar discovery of red radio
galaxies in the LBDS sample by Dunlop and collaborators (Dunlop et
al. 1996; Spinrad et al. 1997; Dunlop 1999). They found two red radio
galaxies with redshifts $z=1.43$ and $z=1.55$ in the 53W field of the
LBDS survey which has a flux-limit of 2 mJy at 1.4 GHz. Peacock et
al. (1998) determined the co-moving space density of these objects to
be $N > 10^{-6.8} {\rm Mpc}^{-3}$ (adjusted for $H_0=50$ km s$^{-1}$
Mpc$^{-1}$ and assuming $\Omega_{\mathrm M}=1$, $\Omega_ \Lambda=0$
initially). Applying a similar process to the red 7CRS galaxies (of
which at least 6 at $1<z<2$ have $R-K>5.5$ - see Fig. \ref{fig:rkz}),
we find a co-moving density of $N = 10^{-7.8} {\rm Mpc}^{-3}$. This is
a factor of 10 lower than the density in the LBDS survey, which is
roughly as expected since the fainter flux limit of the LBDS survey
gives a surface density of radio sources $\approx 20$ times higher
than in the 7CRS.

This hints that there is little correlation between the colours of
$z\sim 1.5$ galaxies and their radio luminosity, so long as one is
well below the extreme radio luminosities of the 3C sample. For
comparison, the typical radio luminosities of $z=1.5$ galaxies in
these samples are $\log_{10} L_{151} = 28$ (3CRR), 26.7 (7CRS), 25.5
(LBDS) ${\rm W Hz}^{-1} {\rm sr}^{-1}$. To get an estimate of the
total space density of radio-loud EROs we integrate the radio
luminosity function of Willott et al. (2000b) down over 2 orders of
magnitude fainter than the 7CRS to $\log_{10} L_{151} = 24$. Radio
sources with luminosities greater than or equal to this value are
almost always due to active nuclei (for comparison M87 has $\log_{10}
L_{151} =24.4 ~{\rm W Hz}^{-1} {\rm sr}^{-1}$). If one now assumes a
similar fraction of high-$z$ radio galaxies at these luminosities are
red as in the 7CRS, the number density of very red radio galaxies
would be $N =10^{-6.0}~ {\rm Mpc}^{-3}$. Note that this calculation
does depend upon whether the form of the RLF of Willott et al. (2000b)
can be reliably extrapolated this far.  This RLF is consistent with
the LBDS data, so any errors in the further extrapolation are likely
to be minimal. It is worth recalling that these calculations are based
on the few radio sources which appear to be extremely red. If the half
of radio galaxies which are blue, are blue because of non-stellar
emission, then their host galaxies (and indeed those of the quasars
which exist in comparable numbers to radio galaxies) may also be very
similar to the EROs. Hence all the number densities quoted above
should be regarded as lower limits which could increase by a factor of
up to $\sim 4$ if all sources contributing to the RLF at these
redshifts (i.e. red radio galaxies, blue radio galaxies and quasars)
have underlying evolved host galaxies.

\subsection{Relationship to field EROs}

The surface density of field EROs with $K < 19.0$ is $0.04$
arcmin$^{-2}$ (Daddi et al. 2000a; Thompson et al. 1999). A simple
calculation (making the crude assumption that these sources lie
randomly distributed in the redshift range $1<z<2$ -- see
e.g. Thompson et al. 1999) gives a volume density of near-IR selected
EROs of $N =10^{-4.5} {\rm Mpc}^{-3}$. Note that the magnitude limit
of this sample is similar to the $K$ magnitudes of the faintest 7CRS
EROs, so we are comparing similarly luminous galaxies.

In the previous section we found that the number density of radio-loud
($L_{151} > 10^{24} ~{\rm W Hz}^{-1} {\rm sr}^{-1}$) EROs is likely to
be at least $10^{-6.0} {\rm Mpc}^{-3}$. This is a factor of 30 lower
than the density of near-IR selected EROs. Hence we find that a small
fraction ($\sim 3$ \%) of field EROs are likely to be hosting
radio-loud AGN (these radio luminosities are much higher than are
achievable by a typical ultraluminous starburst galaxy and therefore
indicate powering by an active nucleus). This hypothesis is directly
testable with a deep VLA survey of a large sample of near-IR selected
EROs.

However, it is well-known that luminous radio sources have limited
lifetimes ($\sim 10^8$ years) which are much smaller than the Hubble
time. The time elapsed between $z=2$ and $z=1$ is approximately 2 Gyr
(for $\Omega_{\mathrm M}=1$, $\Omega_ \Lambda=0$; the time elapsed is
3.5 Gyr for $\Omega_{\mathrm M}=0.3$, $\Omega_
\Lambda=0.7$). Therefore if individual radio sources have lifetimes of
only $\sim 10^8$ years, then the number of galaxies undergoing radio
activity during this period would be a factor of 20 greater than that
observed. Hence all of the near-IR selected EROs could plausibly
undergo such a period of radio activity. A caveat to this is that the
typical lifetimes of weak radio sources such as those which would
dominate the ERO population are not well-constrained. Low-luminosity,
typically FRI structure, sources could have much longer lifetimes of
$\sim 1$ Gyr.  In such a case, only $\sim 10$\% of high-$z$ EROs will
undergo a period of radio activity at some point (but we note there is
evidence that the age of the extended emission in the archetype
low-luminosity radio source in M87 is only $\sim 10^8$ yr --- Owen,
Eilek \& Kassim 2000 --- and this may be typical for low luminosity
radio galaxies). It is also possible that radio sources may undergo
recurrent bursts of activity again reducing the fraction of EROs which
have a period of radio activity at $1<z<2$ [there is no observational
test for recurrent activity over these long timescales ($\sim 10^9$
yr), but multiple radio outbursts over much shorter timescales ($\sim
10^7$ yr) have been observed in a few sources --- e.g. Schoenmakers et
al. 2000].

The percentage of near-IR selected EROs which host radio-loud AGN
derived above ($10 - 100$ \%) is subject to many uncertainties, not
least arising from the masses (and hence $K$-band luminosities) of the
host galaxy counterparts being assumed to be similar for both the 7CRS
EROs and for radio galaxies with still lower radio luminosities.
Indeed, the number counts of EROs at $K\sim 19$ are very steep (Daddi
et al. 2000a), i.e. their space density increases sharply with
decreasing $K$-magnitude. However, there is no problem with the
hypothesis that a significant fraction, and plausibly all, EROs go
through an AGN phase at some point during the period corresponding to
$1<z<2$. The hardness of the X-ray background requires that the space
density of optically-obscured quasars exceeds that of
optically-luminous quasars (e.g. Comastri et al. 1995; Wilman \&
Fabian 1999), which in turn are well known to outnumber radio-loud
quasars by at least an order of magnitude. Many of the hard X-ray
sources discovered in Chandra surveys have very red galaxy
counterparts with weak or absent emission lines (Mushotzky et
al. 2000; Crawford et al. 2000). These objects may well be the
radio-quiet analogues of the 7CRS EROs discussed here (such as the ERO
in the field of 5C7.47 which is discussed in the appendix). The hard
X-ray properties of the ERO population will be investigated with
XMM-Newton and Chandra surveys of ERO fields. Such studies will accurately
determine the AGN contribution, but if radio-quiet quasars have
limited lifetimes, they will not detect objects which are undergoing
inactive phases.

A comparison of the space densities of EROs and present-day
ellipticals has been used as a fundamental test of hierarchical galaxy
formation models (e.g. Zepf 1997; Barger et al. 1999). Daddi et
al. (2000b) use the largest ERO sample currently available to show
that the observed numbers of EROs are consistent with simple
luminosity evolution models where most elliptical galaxies formed at
high redshift. They find that current hierarchical models predict
fewer EROs than are observed. It is clear that some fraction of
massive galaxies did form at high redshifts and also that many (and
plausibly all) of these galaxies are likely to undergo AGN activity at
intermediate ($1 \ltsimeq z \ltsimeq 2$) redshifts. It is likely that
the high redshift starburst which formed the bulk of the stars in
these objects was accompanied by an AGN phase during which the massive
black hole was formed (Fabian 1999; Silk \& Rees 1998). At these high
redshifts, radio galaxies appear to have distorted (rest-frame)
optical structures (e.g. van Breugel et al. 1998), presumably due to a
combination of merging sub-clumps and dust obscuration. Also, the high
incidence of SCUBA sub-mm detection of these radio galaxies (Archibald
et al. 2000) shows they possess large dust masses and probably have
high star-formation rates. The interesting implication from this
section is that many massive galaxies then undergo a second bout of
AGN activity at $z \sim 1.5$. This activity is not accompanied by a
significant amount of star-formation, as revealed by the extremely red
colours of the galaxies and, we suspect, their lack of sub-mm
emission.

\section{Conclusions}

\begin{itemize}

\item{We have shown that the optical to near-infrared SEDs of
optically faint 7CRS radio sources are well fit by old or dusty galaxies
in the redshift range $1 \leq z \leq 2$.}

\item{Two galaxies show narrow emission lines in their near-IR spectra
which if identified as H$\alpha$ give redshifts of $z \sim 1.5$
compatible with the SED fitting. The fluxes of these lines and the
lack of lines in other galaxies are as expected from the correlation
between narrow line and radio luminosities for radio sources.}
 
\item{The reason most of these objects did not have redshifts from
optical spectroscopy is that they fall within the redshift range
where there are no bright lines in the optical wavelength region}.

\item{Independent evidence suggests that the red colours of these
galaxies are not caused by dust-reddening, so they are most likely
very old elliptical galaxies which formed at very early epochs. These
objects account for $\approx 1/4$ of the powerful radio source
population at these redshifts. It is possible that the other radio
sources are blue due to non-stellar emission and have underlying
evolved host galaxies, in which case the space density of radio
galaxies with old stellar populations is a factor of 4 greater than
determined here.}

\item{We find that $\sim 3\%$ of near-IR selected EROs are expected to
be radio-loud AGN with radio luminosities $L_{151} \gtsimeq 10^{24}
~{\rm W Hz}^{-1} {\rm sr}^{-1}$. Depending upon the duration of radio
activity it is possible that a large fraction ($10-100$\%) of near-IR
selected EROs go through such a period of radio activity at some point
in their intermediate-redshift ($1 < z <2$) lifetimes.}

\item{We have discovered a radio-quiet extremely red galaxy
($R-K=5.7$) ERO J081539+2446.8 in the field of 5C7.47 with an emission
line in its optical spectrum. This object is best explained as a
radio-quiet analogue of the 7CRS EROs at $z=1.200$ (see appendix). }

\item{A consistent picture emerges that some of the most massive
galaxies formed in violent starbursts with associated AGN activity at
high redshift ($z > 2$) and then underwent a second bout of AGN
activity at intermediate redshifts ($1<z<2$) but with only a small
amount of further star-formation.}

\end{itemize}

\section*{Acknowledgements}

We would like to thank Mark Lacy, Steve Eales, Gary Hill and Julia
Riley for important contributions to the 7C Redshift Survey. Thanks to
Andrew Bunker, Dan Stern, Hy Spinrad, Adam Stanford, Arjun Dey and
Chuck Steidel for obtaining the Keck data. Thanks to Steve Croft for
obtaining the $R$-band image of 5C6.62, Mark Lacy for obtaining the
NOT data, and Matt Jarvis for help with some of the near-IR
spectroscopy. We acknowledge interesting discussions with Lexi
Moustakas and thank the referee Jim Dunlop for a useful referees
report.  We thank the staff at the WHT, INT and UKIRT for technical
support. The William Herschel Telescope and the Isaac Newton Telescope
are operated on the island of La Palma by the Isaac Newton Group in
the Spanish Observatorio del Roque de los Muchachos of the Instituto
de Astrofisica de Canarias. The United Kingdom Infrared Telescope is
operated by the Joint Astronomy Centre on behalf of the U.K. Particle
Physics and Astronomy Research Council. We acknowledge the UKIRT
Service Programme for some of the near-infrared imaging. Some of the
data presented herein were obtained at the W. M.\ Keck Observatory,
which is operated as a scientific partnership among the California
Institute of Technology, the University of California and the National
Aeronautics and Space Administration.  The Observatory was made
possible by the generous financial support of the W. M.\ Keck
Foundation. The VLA is a facility of the National Radio Astronomy
Observatory (NRAO), which is operated by Associated Universities,
Inc. under a cooperative agreement with the National Science
Foundation. CJW thanks PPARC for support.

\appendix

\section{ERO J081539+2446.8 --  A radio-quiet ERO at $z=1.200$}

\begin{figure*}
\vspace{0.5cm} 
\hspace{-0.9cm} 
\epsfxsize=1.0\textwidth \epsfbox{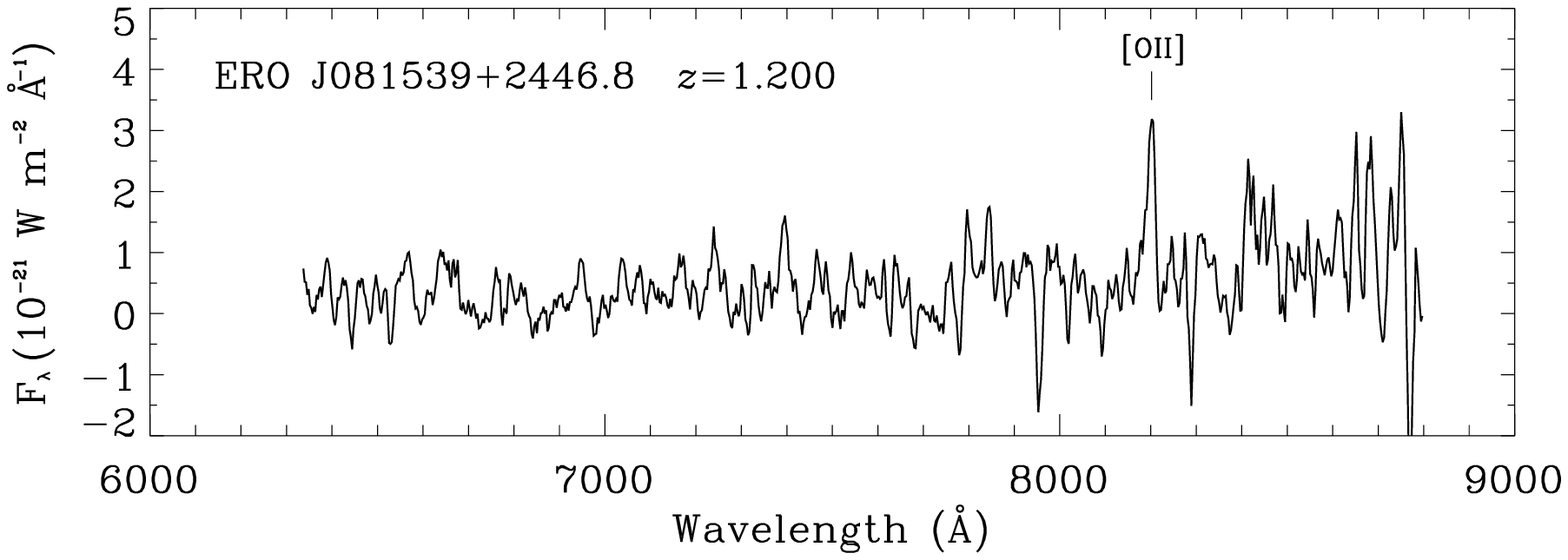}
\vspace{1.5cm} {\caption[junk]{\label{fig:7p47ero} Optical spectrum of
ERO J081239+2456.8 which lies 12.2 arcsec south-east of 5C7.47. The
emission line marked `[OII]'  is the only significant emission line
in the spectrum. This line lies in a gap between the strong sky
emission lines which dominate the background and hence the noise.  The
spectrum shown was extracted within a $2.3 \times 1.5$ arcsec$^2$ aperture
centred on the emission line and has been smoothed with a 14 \AA~
boxcar filter. }}
\end{figure*}

Within 20 arcsec of the radio galaxy 5C7.47 there are two very red
($R-K>5$) galaxies. An excess of EROs in the fields of AGN has been
found in other work (e.g. Yamada et al. 1997; Liu et al. 2000; Cimatti
et al. 2000), although given the sky area covered in total by our
multi-wavelength imaging and the observed surface density of EROs in
near-IR surveys (Daddi et al. 2000a), we would typically expect to
find one serendipitous ERO in our data. These two objects are the
second and third brightest in the $K$-band image shown in
Fig. \ref{fig:ims5c7} and are located 12.2 arsec south-east and 16.5
arcsec north-west of the radio source identification. The object to
the north-west has 3 arcsec aperture magnitudes of $K=19.05 \pm 0.07$
and $R=24.3\pm 0.3$ and is not discussed further. The object to the
south-east [which we designate as ERO J081539+2446.8; co-ordinates 08
15 39.01 +24 46 51.2 (J2000.0)] is slightly redder with 3 arcsec
aperture magnitudes of $K=18.92 \pm 0.06$ and $R=24.6 \pm 0.4$. This
object is detected in the $J$ and $H$ images but with insufficient SNR
for reliable magnitudes to be determined. The object is clearly
resolved at $K$-band with a FWHM of 1.3 arcsec in 1.0 arcsec seeing.

On February 10 2000 we obtained an optical spectrum of 60 minutes
integration of 5C7.47 and this red companion galaxy using the ISIS
long-slit spectrograph on the WHT (for observation and data reduction
details see Willott et al., in prep) . Although no redshift could be
determined for 5C7.47, a single emission feature was detected at the
expected position on the array of ERO J081539+2446.8. This emission
line is located at a wavelength of $8202 \pm 2$ \AA~ and has a FWHM of
18 \AA. The instrumental resolution is 9 \AA~ so this line is clearly
resolved and has a deconvolved velocity width of 570 km s$^{-1}$. The
measured flux in the line is $7\times10^{-20}$ W m$^{-2}$. The
continuum level at the wavelength of the emission line is so weak that
no upper limit of the observed equivalent width can be given. A lower
limit to the observed equivalent width of the emission line is 80 \AA.
Rather fortuitously, the emission line falls within a region of low
sky background in between the strong OH airglow lines which dominate
the $I$-band background spectrum and therefore the SNR of the line is
high (SNR=16). All the other features in the spectrum of Fig.
\ref{fig:7p47ero} are sky subtraction residuals.

Determining redshifts from single emission lines is always an
uncertain process (e.g. Stern et al. 2000) and any other information
available needs to be taken into account. The very red colour of this
object combined with the faint $K$-magnitude suggest $z>0.8$ for
reasons discussed elsewhere in this paper. If this line is Ly$\alpha$
then the derived redshift would be $z=5.75$. There is patchy continuum
in the optical spectrum of ERO J081539+2446.8 down to a wavelength of
at least 7000 \AA. Given that there will be negligible flux below the
Lyman limit, this only puts a firm upper limit on the redshift of
$z<6.68$, so does not rule out the Ly$\alpha$ hypothesis. However,
both the line flux and $K$-magnitude seem too bright for such a
distant galaxy unless there is major non-stellar contamination. In
addition the line does not show an asymmetric profile with blueward
absorption as is typical of Ly$\alpha$ at very high redshifts (Stern
et al. 2000). Another factor arguing against $z=5.75$ is the very red
optical--near-IR SED which would require a large amount of reddening
for a high redshift star-forming galaxy and consequently an even
higher intrinsic emission line flux than is observed.

Hence we think it unlikely that the emission line is Ly$\alpha$. The
two remaining possibilities for single emission lines in an optical
spectrum are MgII $\lambda$2799 and [OII] $\lambda$3727. Considering
MgII initially, the redshift would be 1.93. In this case we would
expect to also detect Ly$\alpha$ emission in the noisy blue end of the
spectrum at 3564 \AA, which is not observed to a $3 \sigma$ limit of
$1.5 \times 10^{-19}$ W m$^{-2}$. This leaves the most likely
explanation that the line is in fact [OII] at a redshift of
$1.200$. Given the observed $K$-magnitude of ERO J081539+2446.8, we
find its luminosity is that of an unevolved $1.6L^{\star}$ galaxy at
$z=1.2$ (for $\Omega_{\rm M}=1$, $\Omega_ \Lambda=0$; for
$\Omega_{\mathrm M} =0.3$, $\Omega_ \Lambda=0.7$ the luminosity is
$3L^{\star}$). This redshift is similar to those of EROs which have
been discovered in near-IR surveys and have redshifts constrained by
the shape of their continua in low-resolution near-IR spectroscopy
(e.g. Cimatti et al. 1999) and also those found in front of the quasar
QSO 1213-0017 (Liu et al. 2000).

The [OII] luminosity for the measured flux at $z=1.2$ is $6 \times
10^{34}$ W ($\Omega_{\rm M}=1$). If this [OII] emission is from a
starburst with extinction similar to local spirals, this implies a
star-formation rate of $30~ {\rm M_{\sun} yr} ^{-1}$ (Kennicutt 1992),
but for a more heavily obscured starburst the actual star-formation
rate would be much higher than this. The extremely red starburst
galaxy HR 10 has a star-formation rate implied by its sub-mm detection
a factor of $50-100$ times greater than inferred from its [OII]
luminosity (Dey et al. 1999). However, it is possible that the
observed emission line is actually powered by an AGN and not
star-formation. We have fit the emission line with the [OII] 3726/3729
doublet to get a more accurate measure of the velocity of the emission
line gas. De-blending the observed line with two components of equal
flux and width, we find the best-fit width for each line is 14 \AA~
(510 km s$^{-1}$). When deconvolved from the instrumental resolution
we find a velocity width of 400 km s$^{-1}$. This is significantly
greater than rotational velocities in massive galaxies at $z \sim 1$
(Vogt et al. 1996) and even high redshift starburst galaxies such as
SMM J14011+0252 (Ivison et al. 2000) and is closer to the widths of
lines in AGN.

ERO J081539+2446.8 is not detected in our 8.4 GHz map of 5C7.47
(Blundell et al. in prep.) or in the FIRST survey at 1.4 GHz
(Becker, White \& Helfand 1995). The $3 \sigma$ limiting flux-density
for the ERO at 1.4 GHz is 0.5 mJy, giving a luminosity limit of
$L_{1.4}<3 \times 10^{23} ~{\rm W Hz}^{-1} {\rm sr}^{-1}$ (assuming a
spectral index of 0.7).  Extrapolating to 151 MHz with this spectral
index gives a limit of $L_{151}<1.3 \times 10^{24} ~{\rm W Hz}^{-1}
{\rm sr}^{-1}$. This upper limit is just at the dividing luminosity
between a radio-loud AGN and extreme starbursts. Given the large
velocity width of the [OII] line we suspect that this object is a
narrow-line AGN, i.e. a quasar with an obscured nuclear region, and
may be similar to the 7CRS EROs in this paper, but radio-quiet. We are
aware of only one other ERO with AGN emission lines - object R7 of Liu
et al. (2000) which is at $z=1.319$. There is no emission in the
spectrum of 5C7.47 at 8202 \AA~ so it seems that the radio source is
not at the same redshift as ERO J081539+2446.8.


\begin{thebibliography}{99}

\bibitem{117} Archibald E.N., Dunlop J.S., Hughes D.H., Rawlings S.,
Eales S.A., Ivison R.J., 2000, MNRAS, in press, astro-ph/0002083

\bibitem{119} Barger A.J., Cowie L.L., Trentham N., Fulton E., Hu
E.M., Songaila A., Hall D., 1999, AJ, 117, 102

\bibitem{318}
Becker R.H., White R.L., Helfand D.J., 1995, ApJ, 450, 559

\bibitem{217} Best P.N., Longair M.S., R\"ottgering H.J.A., 1997,
MNRAS, 292, 758 

\bibitem{218} Best P.N., Longair M.S., R\"ottgering H.J.A., 1998,
MNRAS, 295, 549

\bibitem{245} Blundell K.M., Rawlings S., 1999, Nature, 339, 330

\bibitem{219} Blundell K.M., Rawlings S., Willott C.J., 1999, AJ, 117,
677

\bibitem{357} Bruzual G.A., Magris G.C., 1997, in The Ultraviolet
Universe at Low and High Redshift, AIP vol. 408 (AIP:Woodbury), 291

\bibitem{157} Calzetti D., 1997, AJ, 113, 162 

\bibitem{385} Chambers K.C., Charlot S., 1990, ApJ, 348L, 1

\bibitem{154} Chambers K.C., McCarthy P.J., 1990, 354L, 9

\bibitem{383} Chambers K.C., Miley G., van Breugel W., 1987, Nature,
329, 609

\bibitem{283} Charlot S., Fall S.M., 1993, ApJ, 415, 580

\bibitem{243} Cimatti A., Dey A., van Breugel W., Hurt T., Antonucci
R., 1997, ApJ, 476, 677

\bibitem{288} Cimatti A., Daddi E., di Serego Alighieri S., Pozzetti
L., Mannucci F., Renzini A., Oliva E., Zamorani G., Andreani P.,
R\"ottgering H.J.A., 1999, A\&A, 352L, 45

\bibitem{422} Cimatti A., Villani D., Pozzetti L., di Serego Alighieri
S., 2000, MNRAS, 318, 453

\bibitem{429} Comastri A., Setti G., Zamorani G., Hasinger G., 1995,
A\&A, 296, 1

\bibitem{188} Crawford C.S., Fabian A.C., Gandhi P., Wilman R.J.,
Johnstone R.M., 2000, MNRAS, submitted, astro-ph/0005242

\bibitem{420} Daddi E., Cimatti A., Pozzetti L., Hoekstra H.,
R\"ottgering H.J.A., Renzini A., Zamorani G., Mannucci F., 2000a, A\&A,
361, 535

\bibitem{421} Daddi E., Cimatti A., Renzini A., 2000b, A\&A, in press,
astro-ph/0010093

\bibitem{389} Dey A., van Breugel W., Vacca W., Antonucci R., 1997,
ApJ, 490, 698


\bibitem{323} Dey A., Graham J.R., Ivison R.J., Smail I., Wright G.S.,
Liu M.C., 1999, ApJ, 519, 610

\bibitem{145} di Serego Alighieri S., Fosbury R.A.E., Tadhunter C.N.,
Quinn P.J., 1989, Nature, 341, 307

\bibitem{147} Dunlop J.S., 1999, in The Most Distant Radio Galaxies,
ed. P.N. Best, H.J.A. R\"ottgering, M.D. Lehnert, (KNAW Colloq.;
Dordrecht: Kluwer), 14

\bibitem{138} Dunlop J.S., Peacock J.A., 1990, MNRAS, 247, 19

\bibitem{137} Dunlop J.S., Peacock J.A., 1993, MNRAS, 263, 936

\bibitem{327} Dunlop J.S., Hughes D.H., Rawlings S., Eales S.A., Ward
M.J., 1994, Nature, 370, 347

\bibitem{237} Dunlop J.S., Peacock J.A., Spinrad H., Dey A., Jimenez
R., Stern D., Windhorst R., 1996, Nature, 381, 581

\bibitem{425} Eales S.A., 1992, ApJ, 397, 49 

\bibitem{368} Eales S.A., Rawlings S., 1993, 411, 67

\bibitem{367} Eales S.A., Rawlings S., Dickinson M., Spinrad H., Hill
G.J., Lacy M., 1993, ApJ, 409, 578

\bibitem{175} Eales S.A., Rawlings S., Law-Green J.D.B., Cotter G.,
Lacy M., 1997, MNRAS, 291, 593

\bibitem{176} Efstathiou G., Rees M.J., 1988, MNRAS, 230P, 5

\bibitem{276} Elston R., Rieke M.J., Rieke G.H., 1988, ApJ, 331, L77

\bibitem{278} Elston R., Rieke G.H., Rieke M.J., 1989, ApJ, 341, 80

\bibitem{121} Fabian A.C., 1999, MNRAS, 308L, 39

\bibitem{252} Francis P.J., Hewett P.C., Foltz C.B., Chaffee F.H.,
Weymann R.J., Morris S.L., 1991, ApJ, 373, 465

\bibitem{424} Gardner J.P., Sharples R.M., Frenk C.S., Carrasco B.E.,
1997, ApJ, 480, L99

\bibitem{312} Genzel R., Lutz D., Tacconi L., 1998, Nature, 395, 859

\bibitem{279} Hu E.M., Ridgway S.E., 1994, AJ, 107, 1303

\bibitem{403} Ivison R.J., Smail I., Barger A., Kneib J.-P., Blain
A.W., Owen F., Kerr T., Cowie L., 2000, MNRAS, 315, 209

\bibitem{428} Kennicutt R.C., 1992, ApJ, 388, 310

\bibitem{354} Lacy M., Bunker A.J., Ridgway S.E., 2000, AJ, 120, 68

\bibitem{352} Lacy M., Rawlings S., Eales S.A., Dunlop J.S., 1995,
MNRAS, 273, 821

\bibitem{351} Lacy M., Ridgway S.E., Wold M., Lilje P.B., Rawlings S.,
1999a, MNRAS, 307, 420

\bibitem{350} Lacy M., Rawlings S., Hill G.J., Bunker A.J., Ridgway
S.E., Stern D., 1999b, MNRAS, 308, 1096

\bibitem{309} Laing R.A., Riley J.M., Longair M.S., 1983, MNRAS, 204,
151

\bibitem{310} Lilly S.J., Longair M.S., 1984, MNRAS, 211, 833

\bibitem{412} Liu M.C., Dey A., Graham J.R., Bundy K.A., Steidel C.C.,
Adelberger K.L., Dickinson M.E., 2000, AJ, 119, 2556

\bibitem{134} McCarthy P.J., van Breugel W., Spinrad H., Djorgovski
S., 1987, ApJ, 321, L29

\bibitem{418} Moriondo G., Cimatti A., Daddi E., 2000, A\&A, in press,
astro-ph/0010335

\bibitem{242} Mushotzky R.F., Cowie L.L., Barger A., Arnaud K.A.,
2000, Nature, 404, 459

\bibitem{442} Nolan L.A., Dunlop J.S., Jimenez R., 2000, MNRAS, in
press, astro-ph/0004325

\bibitem{248} Owen F.N., Eilek J.A., Kassim N.E., 2000, ApJ, in press,
astro-ph/0006150

\bibitem{393} Peacock J.A., Jimenez R., Dunlop J.S., Waddington I.,
Spinrad H., Stern D., Dey A., Windhorst R., 1998, MNRAS, 296, 1089
\bibitem{303} Pearson T.J., Kus A.J., 1978, MNRAS, 182, 273
\bibitem{267} Rawlings S., Saunders R., 1991, Nature, 349, 138
\bibitem{355} Rawlings S., Eales S.A., Lacy M., 2000, MNRAS, in press, 
astro-ph/0010445
\bibitem{427} Schoenmakers A.P., de Bruyn A.G., R\"ottgering H.J.A.,
van der Laan H., Kaiser C.R., 2000, MNRAS, 315, 371
\bibitem{229} Silk J., Rees M.J., 1998, A\&A, 331L, 1
\bibitem{423} Simpson C., Rawlings S., Lacy M., 1999, MNRAS, 306, 828
\bibitem{203} Smail I., Ivison R.J., Kneib J.-P., Cowie L.L., Blain
A.W., Barger A.J., Owen F.N., Morrison G., 1999, MNRAS, 308, 1061
\bibitem{257} Soifer B.T., Matthews K., Neugebauer G., Armus L., Cohen
J.G., Persson S.E., Smail I., 1999, AJ, 118, 2065
\bibitem{181} Spinrad H., Dey A., Stern D., Dunlop J., Peacock J.,
Jimenez R., Windhorst R., 1997, ApJ, 484, 58
\bibitem{341} Stern D., Bunker A., Spinrad H., Dey A., 2000, ApJ, 537, 73
\bibitem{441} Stiavelli M., Treu T., 2000, To appear in the
proceedings of the conference ``Galaxy Disks and Disk Galaxies'', ASP
Conf. series, eds. Funes and Corsini, astro-ph/0010100
\bibitem{191} Stiavelli M., Treu T., Carollo C.M., Rosati P., Viezzer
R., Casertano S., Dickinson M., Ferguson H., Fruchter A., Madau P.,
Martin C., Teplitz H., 1999, A\&A, 343, L25
\bibitem{353} Stockton A., Kellogg M., Ridgway S.E., 1995, ApJ, 443L,
69
\bibitem{369} Storrie-Lombardi L.J., McMahon R.G., Irwin M.J., 1996,
MNRAS, 283, 79
\bibitem{146} Tadhunter C.N., Scarrott S.M., Draper P., Rolph C.,
1992, MNRAS, 256, 53
\bibitem{196} Thompson D., Beckwith S.V.W., Fockenbrock R., Fried J.,
Hippelein H., Huang J.-S., von Kuhlmann B., Leinert C., Meisenheimer
K., Phleps S., R\"oser H.-J., Thommes E., Wolf C., 1999, ApJ, 523,
100
\bibitem{268} van Breugel W.J.M., Stanford A.J., Spinrad H., Stern D.,
 Graham J.R, 1998, ApJ, 502, 614 
\bibitem{413} Vogt N.P., Forbes D.A., Phillips A.C., Gronwall C., 
Faber S.M., Illingworth G.D., Koo D.C., 1996, ApJ, 465, L15
\bibitem{414} West M.J., 1991, ApJ, 379, 19 
\bibitem{13}  Willott C.J., Rawlings S., Blundell K.M., Lacy M., 1998,
MNRAS, 300, 625
\bibitem{14}  Willott C.J., Rawlings S., Blundell K.M., Lacy M., 1999,
MNRAS, 309, 1017
\bibitem{15}  Willott C.J., Rawlings S., Blundell K.M., Lacy M., 2000a,
MNRAS, 316, 449
\bibitem{17} Willott C.J., Rawlings S., Blundell K.M., Lacy M., Eales
S.A., 2000b, MNRAS, in press, astro-ph/0010419
\bibitem{430} Wilman R.J., Fabian A.C., 1999, MNRAS, 309, 862
\bibitem{376} Worthey G., 1994, ApJS, 95, 107 
\bibitem{478} Yamada T., Tanaka I., Aragon-Salamanca A., Kodama T., Ohta K., Arimoto N., 1997, ApJ, 487, L125
\bibitem{378} Yi S., Brown T., Heap S., Hubeny I., Landsman W., Lanz
T., Sweigart A., 2000, ApJ, 533, 670
\bibitem{426} Zepf S.E., 1997, Nature, 390, 377 


\end{thebibliography}
\end{document}